\documentclass[useAMS,usenatbib,usegraphicx]{mn2e}

\title[Orbits of Open Clusters]{The orbits of open clusters in the Galaxy}
\author[Wu et al.]{Zhen-Yu Wu$^{1}$\thanks{E-mail: zywu@bao.ac.cn}, Xu Zhou$^{1}$, Jun Ma$^{1}$ and Cui-Hua Du$^{2,\,1}$\\
$^{1}$National Astronomical Observatories, Chinese Academy of Sciences, 20A Datun Road, Beijing 100012, China\\
$^{2}$College of Physical Sciences, Graduate University of the Chinese Academy of Sciences, Beijing 100049, China}

\begin{document}
\date{}
\pubyear{2009} \maketitle

\begin{abstract}
We present and analyze kinematics and orbits for a sample of 488 open clusters in the Galaxy. The velocity ellipsoid for our present sample is derived as ($\sigma_{U}$, $\sigma_{V}$, $\sigma_{W})$=$(28.7$, $15.8$, $11.0$) km s$^{-1}$ which represents a young thin disc population. We also confirm that the velocity dispersions increase with the age of cluster subsample. The orbits of open clusters are calculated with three Galactic gravitational potential models. The errors of orbital parameters are also calculated considering the intrinsic variation of the orbital parameters and the effects of observational uncertainties. The observational uncertainties dominate the errors of derived orbital parameters. The vertical motions of clusters calculated using different Galactic disc models are rather different.  The observed radial metallicity gradient of clusters is derived with a slope of $b=-0.070\pm0.011$ dex kpc$^{-1}$. The radial metallicity gradient of clusters based on their apogalactic distances is also derived with a slope of $b=-0.082\pm0.014$ dex kpc$^{-1}$. The distribution of derived orbital eccentricities for open clusters is very similar to the one derived for the field population of dwarfs and giants in the thin disc.
\end{abstract}


\section{Introduction}
Open clusters (OCs) have long been used as important tools in the study of the Galactic disc. The young clusters have been used to determine spiral arm structure, and to map the rotation curve of the Galaxy. The old clusters are excellent tracers of the structure, kinematics, and chemistry of the Galactic disc \citep{fr95}. In principle, basic parameters like distance, age, and metallicity can be determined for a cluster more accurately than for a field star. Therefore, OCs are better tracers of large scale properties of the Galactic disc population than field stars \citep{pi06}.

About 20 years ago, the catalogue of open cluster data compiled by Lyng\r{a} \citep{ly87a} was used to derive the radial gradients and other structural properties of the Galactic disc \citep*{ly82, j88}. In recent years, the available data on open clusters have increased very quickly and the basic parameters of these clusters have also considerably improved.  \citet[hereafter DAML]{di02} compiled the new catalogue of optically visible open clusters and candidates including 1.5 times more clusters than the catalogue of \citet{ly87a}. \citet[hereafter K05]{kh05} presented a catalogue of astrophysical data for 520 OCs, with data derived from the All-Sky Compiled Catalogue of 2.5 Million Stars \citep[hereafter ASCC-2.5] {kh01}. Using these data, the properties of the Galactic disc such as the scale height, the scale length, and the metallicity distribution of the disc, were derived \citep*{ch03, bo06, pi06}.

\citet{ly87b} used 106 OCs with available data on positions, distances, and radial velocities to analyze the local kinematics. They found that the dispersion of the radial velocity increases with age: for the old clusters it is about twice that of young clusters. \citet*[hereafter BKO]{ba87} calculated the Galactic orbits for 69 OCs based on carefully selected data on distances, absolute proper motions, and radial velocities. \citet*{ke73} calculated the orbits of NGC 188 and M67. In addition, \citet{as88} calculated the orbits for NGC 188, M67, and NGC 2420, \citet[hereafter C94]{ca94} expanded the sample with 5 classic clusters, and \citet{fi95} calculated the orbits for a total of 7 old clusters. \citet*{so00} calculated the orbit for NGC 2355, \citet{de02} calculated the orbits for NGC 1912 and NGC 1907, and \citet{wu02} calculated the orbit for M48.

In recent years, \citet{di05} determined the rotation velocity of the spiral pattern of the Galaxy by studying the birthplaces of OCs in the Galactic disc as a function of their ages. The birthplaces of these clusters were determined by assuming \textbf{that their orbits were circular.} Using 148 OCs within the projected distance onto the Galactic plane $d_{xy}\le 0.85$ kpc, \citet{pi06} derived the Solar motion and the velocity ellipsoid of OCs. They also calculated the Galactic orbits of these clusters and presented the mean parameters of their orbits. Based on the spatial and velocity distributions of OCs, \citet{pi06} also identified the existence of four open cluster complexes of different ages \citep{kh06, ro07}, \textbf{which verifed the nature of clustering in OCs in the Solar neighbourhood identified in previous studies\citep*{ef78, ei88, ba89}.} \citet*{le08} used 374 OCs taken from the 2.7 version of DAML catalogue to measure the epicycle frequency $k$ in the Galactic disc. They also calculated the orbits of these OCs and discussed the distribution of initial velocities of these clusters.

The orbital motions of OCs are important not only for our understanding of the dynamical evolution of OCs in the Galaxy \citep*{fr99, be01}, but also for investigating their effect on the time-evolution of the abundance gradient in the Galactic disc (C94). The main aim of this paper is to calculate the orbits of OCs with the improved data for an enlarged sample of 488 OCs and to discuss the kinematical properties of these clusters.

We present the collection of data in Section 2. In Section 3, we analyze the statistical properties of the sample, and especially their age-velocity dispersion relations. We present the orbital parameters and their associated uncertainties calculated in a given Galactic potential model in Section 4, followed by Section 5, which presents the differences of orbital parameters due to different Galactic potential models. In Section 6, we compare our results with the orbital parameters calculated for field stars and globular clusters, and discuss the effect of the orbital motions of OCs for the radial abundance gradient derived for these clusters, we also compare the orbital parameters with those derived by previous studies. Our conclusions are given in Section 7.

\section{The sample}
\subsection{The DAML catalogue}
We chose the DAML catalogue as the main source of the fundamental parameters for the OCs. This catalogue uses the WEBDA database\footnote{http://obswww.unige.ch/webda} and previous catalogues of \citet{ly87a} as a starting point. Kinematical, and metallicity data of the new objects when available are inserted \citep{di02}. They also made use of the Simbad database\footnote{http://simbad.u-strasbg.fr/simbad/} and of the literature to find data on the clusters or on individual stars of the clusters, to obtain the averaged values of radial velocities and proper motions. This catalogue is regularly updated, and the latest version is available from its website\footnote{http://www.astro.iag.usp.br/\~ \,wilton/.}. The present 2.9 version (13/Apr/2008) of DAML catalogue contains 1776 objects, of which 936 have published distances, ages, and reddening values, 890 have published proper motions and 447 have radial velocities. 869 clusters with distances and proper motions data are taken from the DAML catalogue as the initial sample.
\subsection{Distances}
Most of distances and ages listed in DAML are taken from WEBDA, and  are updated with new data from literature \citep{di02}. The distances estimated by \citet*{ba00} based on Hipparcos parallaxes \citep{esa} of member stars are also adopted by DAML. A comparison of the Hipparcos parallaxes with photometric distances shows good agreement \citep{ba00}. The distances listed in DAML are taken from different authors using different observational techniques and reduction methods, and no errors for this parameter are listed in DAML, the non-uniformity and uncertainties in the adopted distance data should be estimated.

\citet{pa06} studied the accuracy of available parameters such as the age, reddening, and distance for OCs by using the independently derived values published in the literature. They used a sample of 395 clusters in their statistical analysis. They found that, for about 80\% clusters in their sample, the error of distance is less than 20\%. They compared their results with the data of DAML catalogue, and pointed out that the distances listed in DAML catalogue have the same error distributions. In our present study, 20\% relative errors in distances are adopted for OCs in our sample.

\citet{vl07a} re-reduced the raw data of Hipparcos mission, and the new reduction provides an improvement by a factor 2.2 compared to the catalogue published in 1997 \citep{vl07b}. We calculate the difference between the distances derived by \citet{vl07a} using the re-reduced Hipparcos data and that listed in DAML catalogue for a sample of 17 OCs in common. We get a difference of $6.2 \pm 1.6\%$ between these two data sets, which is less than our adopted $20\%$ error for distance parameters and is considered in the following orbit calculation.
\subsection{Absolute proper motions}
The absolute proper motion data of OCs listed in DAML are adopted in the present study. The proper motion data are also compiled from different authors, but they are all based on the Hipparcos system. In our final sample, proper motions of $15\%$ clusters are derived from the Hipparcos catalogue \citep{esa}, $47\%$ proper motions are derived from the Tycho2 catalogue \citep{ho00}, $36\%$ proper motions are derived from the ASCC-2.5 catalogue \citep{kh01}, and only $2\%$ proper motions are derived from the UCAC2 catalogue \citep{za04}.

The Hipparcos catalogue is considered as the realization of the ICRS at optical wavelengths. The systematic error of proper motions in the Hipparcos catalogue with respect to ICRS is estimated to be $0.25 \,\textrm{mas yr}^{-1}$ \citep{esa, ko97}. \citet*{pl98} searched the Hipparcos catalogue and found 9 new OCs and derived the proper motions for these clusters. \citet{ba00} determined mean proper motions of 205 OCs from their member stars found in the Hipparcos catalogue. In our final sample, the proper motions derived from Hipparcos catalogue are all taken from the results of \citet{pl98} and \citet{ba00}.

The Tycho2 catalogue \citep{ho00} presents very precise proper motions with random errors between 1 and 3 $\textrm{mas yr}^{-1}$ in the Hipparcos system. There are no significant systematic differences between the proper motions of these two catalogues \citep*{ur00}. \citet*{di01,di02p} determined the mean absolute proper motions of 206 OCs from the data in Tycho2 catalogue. \citet{al03} found 11 new OCs candidates in Tycho2 catalogue and determined the mean proper motions for these clusters. \citet{lo03} determined the mean proper motions for 167 OCs based on the kinematic and photometric data in Tycho2 catalogue. \citet*{di01,di02p} compared their results with those derived by \citet{ba00} based on Hipparcos catalogue, and found that the mean difference in the proper motions is less than 1 mas yr$^{-1}$. \citet{lo03} also compared their results with those derived by \citet*{di01,di02p} and found that the difference is $4 \pm 5\%$. In our final sample, the mean proper motions derived from Tycho2 catalogue are taken from above mentioned studies.

The ASCC-2.5 catalogue is based on large, modern, high-precision catalogues of the Hipparcos-Tycho family, including the Tycho2 catalogue, and provides the most complete all-sky catalogue of stars having uniform high precision astrometric and photometric data down to $V \sim 14$ mag \citep{kh01}. In our final sample, the mean proper motions derived from ASCC-2.5 catalogue by \citet{kh03, kh05} are adopted. \citet{kh03,kh05} compared their results with those derived by \citet{ba00} and \citet*{di01,di02p} and found that their results agree quite well with previous studies.

The UCAC2 catalogue presents proper motions in the Hipparcos system with nominal errors of 1 to 3 $\textrm{mas yr}^{-1}$ for stars up to $V \sim 12$ mag and about 4 to 7 $\textrm{mas yr}^{-1}$ for fainter stars up to $V \sim 16$ mag. The systematic errors of the proper motions in UCAC2 are in the range 0.5 to 1.0 $\textrm{mas yr}^{-1}$ \citep{za04}. \citet{di06} determined the mean proper motions for 428 OCs from the UCAC2 catalogue. They compared their results with those derived from Hipparcos catalogue \citep{ba00}, Tycho2 catalogue \citep*{di01,di02p, lo03} and ASCC-2.5 catalogue \citep{kh03}, and found that there is no statistical distinction between the compared mean proper motions of OCs in these catalogues.

We also calculate the difference of the mean proper motions derived by \citet{vl07a} using the re-reduced Hipparcos data and that adopted in our final sample taken from the DAML catalogue for the 17 OCs in common. We get a difference of $0.35 \pm 0.22$ mas yr$^{-1}$ in $\mu_{\alpha}\cos\delta$ and a difference of $0.46 \pm 0.43$ in $\mu_{\delta}$. The differences are within the range of  the errors for the mean proper motions listed in the DAML catalogue.
\subsection{Radial velocities}
There are 431 OCs whose distances, proper motions, and radial velocities are available from the DAML catalogue. Most radial velocities are taken from the catalogues compiled by \citet{kh05, kh07}. \citet{kh05} cross-identified the ASCC-2.5 catalogue with the General Catalogue of Radial Velocities of \citet{bb00} and derived radial velocities for 290 OCs based on their membership determination. They found that the mean difference between their results and those in the published literature for common clusters is $0.36 \pm 0.88$ km s$^{-1}$. The 363 radial velocities of OCs compiled by \citet{kh07} are derived from the 2nd version of the Catalogue of Radial Velocities of Galactic stars with high precision Astrometric Data (CRVAD-2). The CRVAD-2 is the result of updating and expanding the list of stars with known radial velocities and high precision astrometric and photometric data taken from the ASCC-2.5 catalogue. The mean difference of radial velocities for 177 clusters in common between the CRVAD-2 and the literature  is $0.65 \pm 0.72$ km s$^{-1}$.

\citet{fr08} presented high-precision radial velocities for 71 OCs obtained with multi-object spectrographs. For 25 clusters in their sample, the radial velocities are newly obtained and are not included in DAML. \citet*{me08} derived mean radial velocities for 166 OCs based on observations with the CORAVEL spectrovelocimeters. The radial velocities of 64 OCs derived by \citet{me08} are not listed in DAML. We supplement the new radial velocities derived by \citet{fr08} and \citet{me08} into our final sample. We also update the known radial velocities in DAML with those derived by \citet{me08} due to their higher precision.

In our final sample, 488 OCs with distances, proper motions, and radial velocities are included.
\subsection{Completeness of present sample}
\begin{figure*}
\begin{center}
 \includegraphics[width=160mm]{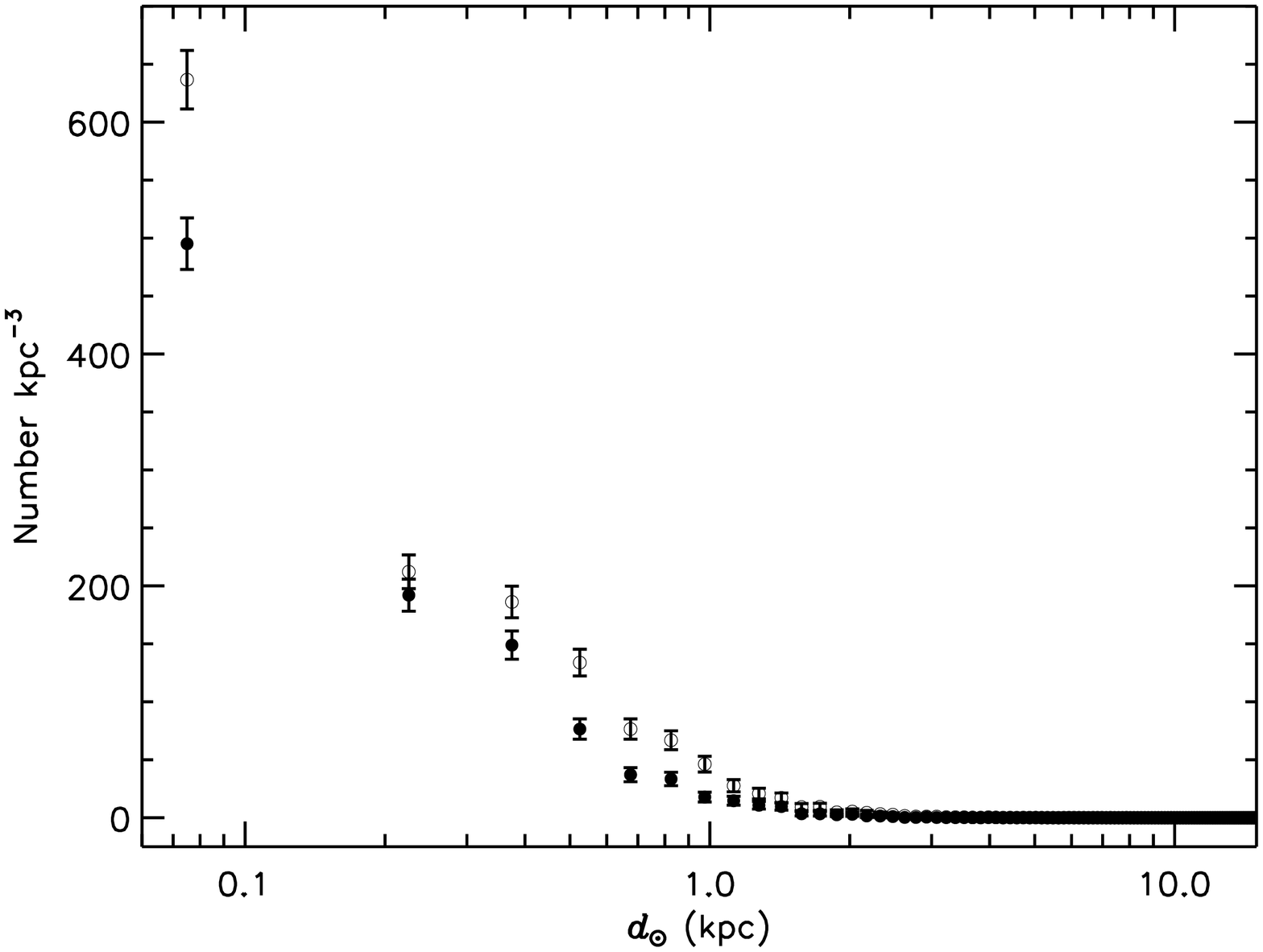}
\caption{The volume density distributions of OCs as a function of their  heliocentric distance $d_{\odot}$. The open circles represent the OCs in DAML catalogue with distance data available, the filled circles represent the OCs in our present sample.}\label{incom}
\end{center}
\end{figure*}

Our present sample is a subsample mainly taken from the DAML catalogue. We plot the volume density of  1082 OCs (open circles) with distance data available in DAML as a function of heliocentric distance $d_{\odot}$ in Fig.~\ref{incom}. In order to estimate the completeness of the OCs in our present sample,  we also plot the volume density of 488 OCs (filled circles) in our present sample in Fig.~\ref{incom}. For clusters with heliocentric distance $d_{\odot} > 2.0$ kpc,  the volume density distributions for these two samples of OCs are  completely consistent. For clusters with heliocentric distance $d_{\odot} < 2.0$ kpc, the volume densities of OCs in our present sample are less than those for OCs in DAML, but the distributions for these two samples are very similar. Fig.~\ref{incom} indicates that our present OCs sample is a representative subsample of currently observed OCs in the Galaxy.

Using a sample of 654 OCs with distance data available, \citet{bo06} simulated the effects of completeness in their OCs sample. They found that a total number of  $\sim 730$ OCs with heliocentric distance $d_{\odot} \le 1.3$ kpc should be observed. Within the same distance range, there are 498 OCs in DAML catalogue and 271 OCs in our present sample, the completeness can be estimated as $68\%$ and $37\%$ for DAML catalogue and for our present sample, respectively. A possible source for the remainder unobserved OCs comes from the very young clusters which are still embedded within giant molecular clouds,  they are heavily obscured and are very difficult to identify \citep{la03}.

\subsection{Initial conditions}
Table~\ref{tb1} lists the adopted positions ($\alpha, \delta$), heliocentric distances $d_{\sun}$, radial velocities $v_{r}$, and absolute proper motions ($\mu_{\alpha}\cos\delta$, $\mu_{\delta}$) of 488 OCs in our final sample. The adopted $20\%$ errors for distances, the errors of absolute proper motions and of radial velocities listed in DAML, \citet{fr08}, and \citet{me08} are also presented in Table~\ref{tb1}. In Table~\ref{tb1}, we also list the ages and metallicities [Fe/H] for each cluster if  available in DAML\@. Ages are currently available for 445 OCs, 109 of which have [Fe/H] values also available.

The initial conditions for orbit calculation are the presently observed positions and velocities of OCs with respect to the galactocentric reference frame. Adopting the solar motion ($U$, $V$, $W$)$_{\sun}$ $=$ (10.0, 5.2, 7.2) km s$^{-1}$ from \citet{de98}, the local standard of rest (LSR) velocities of OCs are determined from the data listed in Table~\ref{tb1}. The LSR velocities are then corrected to the Galactic standard of rest (GSR) by adopting the galactocentric distance of Sun $R_{\sun}=8.0$ kpc \citep{re93} and a rotation velocity of the LSR of $220$ km s$^{-1}$\citep{ke86}. All space coordinates $x$, $y$, $z$, and velocity components $U$, $V$, $W$ refer to a galactocentric righthanded cartesian coordinate system with the $x$ direction directed towards the Galactic anticenter and the $z$ direction directed towards the Galactic north pole \citep{od92}. The $U$, $V$, $W$ velocity components and their errors are calculated with the method of \citet{jo87}. Table~\ref{tb2} lists the initial conditions used to calculate the orbital solution for each cluster in our sample. The errors in velocity components include the errors in absolute proper motions, radial velocities, and distances of OCs listed in Table~\ref{tb1}.

\begin{table*}
\begin{minipage}{175mm}
\caption{The observed data of 488 OCs in our sample. The full version of this table is available in the online version of this article.}\label{tb1} \centering
\begin{tabular}{lcccccccc}
\hline name & $\alpha (2000.0)$ & $\delta (2000.0)$ & $d_{\sun}$ &$v_{r}$& $\mu_{\alpha}\cos\delta$& $\mu_{\delta}$&Age&[Fe/H]\\
 & h m s& \degr\, \arcmin\, \arcsec& (kpc)& (km s$^{-1})$& (mas yr$^{-1})$& (mas yr$^{-1}$)& (Myr)&\\ \hline
Berkeley 59&$00:02:14$&$+67:25:00$&$1.000\pm0.200$&$-12.5\pm
7.1$&$-2.11\pm0.81$&$-1.20\pm0.75$&   6.3&$     $\\
Blanco 1   &$00:04:07$&$-29:50:00$&$0.269\pm0.054$&$  4.1\pm
1.4$&$20.17\pm0.51$&$ 3.00\pm0.51$&  62.5&$ 0.04$\\
Alessi 20  &$00:09:23$&$+58:39:57$&$0.450\pm0.090$&$-11.5\pm 0.0$&$
8.73\pm0.53$&$-3.11\pm0.53$& 166.0&$     $\\
ASCC 1     &$00:09:36$&$+62:40:48$&$4.000\pm0.800$&$-69.7\pm
4.7$&$-2.07\pm0.72$&$ 0.46\pm0.57$& 177.8&$     $\\
Mayer 1    &$00:21:54$&$+61:45:00$&$1.429\pm0.286$&$-20.9\pm
2.0$&$-4.46\pm1.13$&$-6.66\pm0.94$&      &$     $\\
NGC 129    &$00:30:00$&$+60:13:06$&$1.625\pm0.325$&$-39.4\pm
0.5$&$-1.06\pm0.94$&$ 1.60\pm0.94$&  76.9&$     $\\
ASCC 3     &$00:31:09$&$+55:16:48$&$1.700\pm0.340$&$-37.0\pm
0.0$&$-1.92\pm0.61$&$-1.25\pm0.59$&  79.4&$     $\\
NGC 225    &$00:43:39$&$+61:46:30$&$0.657\pm0.131$&$-28.0\pm
0.0$&$-4.95\pm0.76$&$-0.50\pm0.76$& 130.0&$     $\\
NGC 188
&$00:47:28$&$+85:15:18$&$2.047\pm0.409$&$-45.0\pm10.0$&$-1.48\pm1.25$&$-0.56\pm1
.24$&4285.5&$-0.01$\\
IC 1590    &$00:52:49$&$+56:37:42$&$2.940\pm0.588$&$-32.5\pm
6.4$&$-1.36\pm0.23$&$-1.34\pm0.83$&   3.5&$     $\\
\ldots&\ldots&\ldots&\ldots&\ldots&\ldots&\ldots&\ldots&\ldots\\
\ldots&\ldots&\ldots&\ldots&\ldots&\ldots&\ldots&\ldots&\ldots\\
\hline
\end{tabular}
\end{minipage}
\end{table*}

\begin{table*}
\begin{minipage}{175mm}
\caption{The present positions and velocities of 488 OCs in our sample. The full version of this table is available in the online version of this article.}\label{tb2} \centering
\begin{tabular}{lccccccc}
\hline
name & $x$ & $y$ &$z$& $U$&$V$&$W$\\
         &\multicolumn{3}{c}{ (kpc)}&\multicolumn{3}{c}{ (km s$^{-1}$)}\\
\hline Berkeley 59&$-8.471\pm0.094$&$0.878\pm0.176$&$
0.087\pm0.017$&$ 25.3\pm
5.1$&$219.7\pm6.6$&$  2.4\pm0.7$\\
Blanco 1
&$-7.952\pm0.010$&$0.013\pm0.003$&$-0.264\pm0.053$&$-13.4\pm
4.9$&$217.1\pm1.8$&$ -1.7\pm0.7$\\
Alessi 20  &$-8.207\pm0.041$&$0.398\pm0.080$&$-0.030\pm0.006$&$
0.2\pm
3.2$&$206.5\pm1.8$&$ -1.6\pm0.2$\\
ASCC 1     &$-9.887\pm0.377$&$3.527\pm0.705$&$ 0.014\pm0.003$&$
75.8\pm13.9$&$181.3\pm8.4$&$ 21.9\pm0.3$\\
Mayer 1    &$-8.702\pm0.141$&$1.244\pm0.249$&$-0.023\pm0.005$&$
51.2\pm
9.1$&$223.7\pm5.3$&$-33.8\pm0.5$\\
NGC 129    &$-8.818\pm0.164$&$1.402\pm0.280$&$-0.072\pm0.014$&$
35.7\pm
6.4$&$195.3\pm3.9$&$ 21.9\pm0.7$\\
ASCC 3     &$-8.843\pm0.169$&$1.459\pm0.292$&$-0.221\pm0.044$&$
43.0\pm
5.2$&$200.6\pm3.0$&$  3.3\pm0.0$\\
NGC 225    &$-8.348\pm0.069$&$0.557\pm0.111$&$-0.012\pm0.002$&$
38.0\pm
3.3$&$209.6\pm2.1$&$  6.6\pm0.4$\\
NGC 188    &$-9.027\pm0.205$&$1.590\pm0.318$&$ 0.780\pm0.156$&$
43.6\pm11.8$&$199.7\pm1.0$&$-14.7\pm0.8$\\
IC 1590    &$-9.597\pm0.319$&$2.448\pm0.490$&$-0.320\pm0.064$&$
44.6\pm
5.6$&$206.7\pm6.0$&$ -7.9\pm0.1$\\
\ldots&\ldots&\ldots&\ldots&\ldots&\ldots&\ldots\\
\ldots&\ldots&\ldots&\ldots&\ldots&\ldots&\ldots\\
\hline
\end{tabular}
\end{minipage}
\end{table*}

\section{Parameter distributions and kinematics of the sample}
\subsection{Parameter distributions of the present sample}
\begin{figure*}
\begin{center}
 \includegraphics[width=160mm]{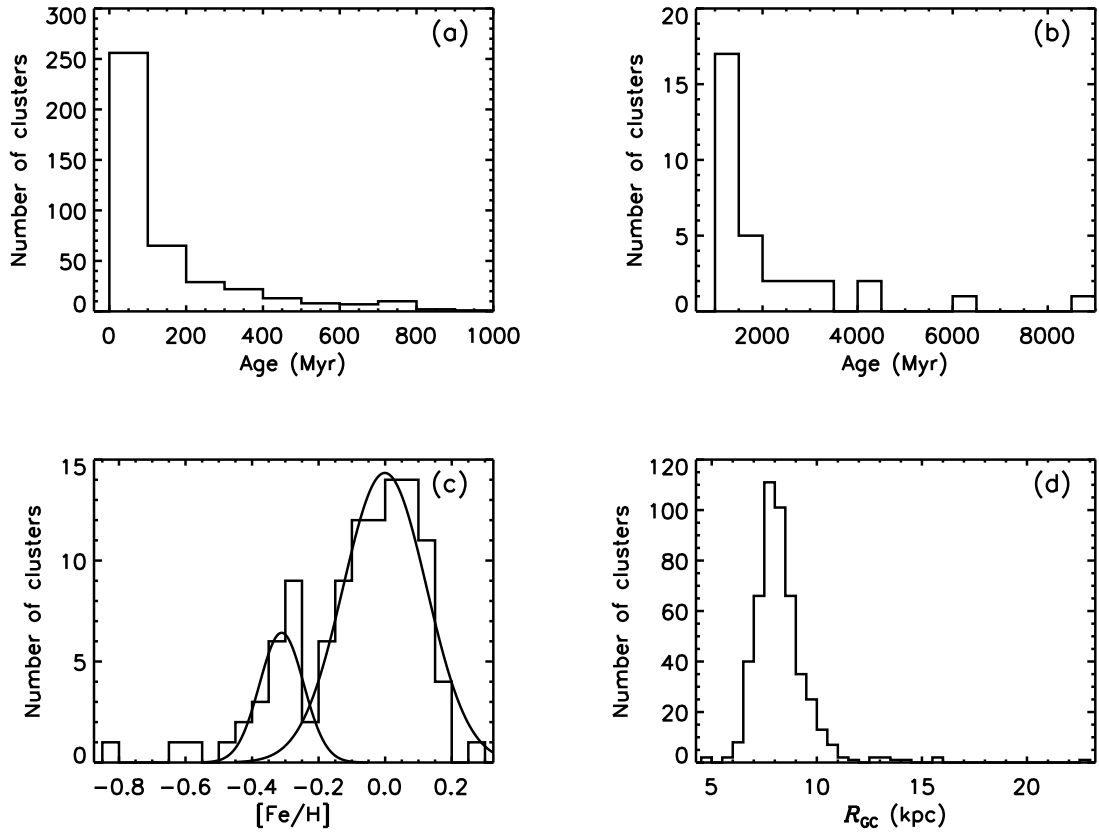}
\caption{Distributions of ages, metallicities [Fe/H], and observed galactocentric distances $R_{\textrm{GC}}$ of OCs in our present sample.}\label{hsam}
\end{center}
\end{figure*}

Fig.~\ref{hsam} shows the distributions of ages, metallicities [Fe/H], and observed galactocentric distances $R_{\textrm{GC}}$ of OCs in our present sample. Due to the large difference among the ages of OCs, the distributions of age are plotted in panels a and b of Fig.~\ref{hsam}. We can see from panels a and b of Fig.~\ref{hsam} that, about 2/3 OCs have ages less than 100 Myr. The oldest cluster has an age of 9.0 Gyr. No clusters in our sample have been found to be in the age interval 0.9 -- 1.0 Gyr. The distribution of metallicities [Fe/H] of OCs is plotted in panel c of Fig.~\ref{hsam}. Except for two clusters with [Fe/H] $< -0.5$, the distribution of [Fe/H] can be fitted by two Gaussian functions. For clusters with $-0.5<$ [Fe/H] $< -0.2$, the [Fe/H] data can be fitted by a Gaussian function with mean $\mu_{\textrm{[Fe/H]}}=-0.31$ and dispersion $\sigma_{\textrm{[Fe/H]}}=0.07$; for clusters with  [Fe/H] $> -0.2$, the best-fitting Gaussian function has mean $\mu_{\textrm{[Fe/H]}}=0.0$ and dispersion $\sigma_{\textrm{[Fe/H]}}=0.13$. It should be noted that the sample of clusters with [Fe/H] data is very incomplete, the distribution of [Fe/H] in panel c of Fig.~\ref{hsam} may not be the true distribution of the metallicities for the complete OCs sample in the Galaxy. From panel d of Fig.~\ref{hsam}, it can be seen that most OCs distribute near the Sun. The minimum and maximum galactocentric distances of OCs in our present sample are $4.6$ and $22.6$ kpc respectively.

\subsection{Age-velocity dispersion relations}
In recent years, many observational efforts have been devoted to constrain the age-velocity dispersion relation of the thin disc. \citet[hereafter N04]{no04} presented new determinations of metallicity, age, kinematics, and Galactic orbits for a complete, magnitude-limited, and kinematically unbiased sample of $\sim 14000$ F and G dwarf stars near the Solar neighborhood. The Hipparcos/Tycho-2 parallaxes and proper motions, together with some 63000 new, accurate radial velocity observations supplemented by a few earlier radial velocities, were used to compute the space velocity components and their dispersions. Ages and their errors were computed from a set of theoretical isochrones by a sophisticated Bayesian technique \citep{jl05}. N04 found that the age-velocity dispersion relations of each space velocity component can be fitted by continuous smooth power laws: $\sigma \propto \textrm{age} ^{k}$, which also evidences the continuous heating of the disc in all directions.

\citet*{ho07} redetermined the basic calibrations used to infer astrophysical parameters for the N04 stars from $uvby$ photometry. Using the improved astrophysical parameters, they  recomputed the ages and age error estimates for the N04 sample. Based on their revised data set, and with substantially higher time resolution than that in the original N04, \citet{ho07} confirmed the conclusion of N04 that the dynamical heating of the thin disc continues throughout its life.

\citet{se07} revisited the Galactic thin disc age-velocity dispersion relation based on the N04 sample, their new result is that a power law is not required by the data of N04, and disc heating models that saturate after $\sim$ 4.5 Gyr are equally consistent with the observations.

\citet[hereafter S08]{so08} presented the parameters of 891 stars, mostly local and distant clump giants, including distances, absolute magnitudes, spatial velocities, galactic orbits, and ages.  Using their distant sample of clump giants, and rejecting stars having a probability higher than 80\% to belong to the thick disc, the Hercules stream, and the halo, S08 found that the velocity dispersions in $V$ and $W$ saturate at $\sim$ 4 Gyr and the dispersion in $U$ increases smoothly with time.

Using radial velocities of 67 clusters within 2 kpc from the Sun, \citet{ly87b} found that the dispersion of the radial velocity increases with age.  Based on proper motions and distances of 148 clusters within  the projected distance onto the Galactic plane $d_{xy} \le 0.85$ kpc from the Sun, \citet{pi06} derived the tangential velocity dispersions  for  clusters with different ages and also found that the dispersions increase with age.

The OCs in our sample with errors in the spatial velocities less than 20 km s$^{-1}$ are used to derive the  velocity dispersion in each velocity component. Those clusters are divided into three age groups: $0 <$ age $\le 500$ Myr, $500<$ age $\le 1000$ Myr,  and $1000<$ age $\le 2000$ Myr. For each age group, the velocity dispersion in each velocity component $U$, $V$, and $W$ is calculated. The number $N$ of clusters in each age group is also listed in Table~\ref{tb3}. Table~\ref{tb3} indicates that the velocity dispersions in the $U$ and $W$ components in the  age group of $500<$ age $\le 1000$ Myr are only marginally larger than those in the age group with age $< 500$ Myr,  but the velocity dispersions in all of the three velocity components in the age group with age $> 1$ Gyr are  larger than those for the clusters with age $< 1$ Gyr. The velocity dispersions in our present OCs sample indicate the continuous dynamical heating of the thin disc.

Using OCs  with errors in the spatial velocities less than 20 km s$^{-1}$ in our present sample,  the velocity dispersions are derived as ($\sigma_{U}$, $\sigma_{V}$, $\sigma_{W})=$(28.7,  15.8,  11.0) km s $^{-1}$.  The derived velocity dispersions are smaller than those for the thin disc clump giants  ($\sigma_{U}$, $\sigma_{V}$, $\sigma_{W})=(41.5$, $26.4$, $22.1)$ km s $^{-1}$ derived by S08.  For their subsample of clump giants with age of 1.5 Gyr, S08 derived the velocity dispersions as ($\sigma_{U}$, $\sigma_{V}$, $\sigma_{W})=(36.2$, $19.9$, $18.7)$ km s $^{-1}$, which are close to our results for the OCs with age $> 1$ Gyr. So, the main reason for the difference between the velocity dispersions for OCs in our present sample and those for the giants of S08 is that in our present OCs sample, most of them are young clusters with age less than 1 Gyr. The giants sample of S08 represents an old thin disc, most of giants in their sample are older than 1 Gyr. Just as we have pointed out in the previous section, the thin disc of the Galaxy is continuously heated, so the velocity dispersions for a young population such as the OCs in our present sample should be less than that corresponding to an older population such as the giants sample of S08. 

Using OCs within  the projected distance onto the Galactic plane $d_{xy} \le 0.85$ kpc from the Sun, \citet{pi06} derived the velocity dispersions as  ($\sigma_{U}$, $\sigma_{V}$, $\sigma_{W})=$(13.86,  8.75,  5.05) km s $^{-1}$, which are smaller than those derived from our present sample. \citet{pi06} used a volume-limited sample of relatively young clusters; in contrast, our present sample includes many older clusters and also objects located further away. Therefore, our larger velocity dispersions reflect the effects of dynamical heating of the thin disc. Using the clusters in our present sample within the projected distance onto the Galactic plane $d_{xy} \le 0.85$ kpc, we derive the velocity dispersions as ($\sigma_{U}$, $\sigma_{V}$, $\sigma_{W})=$(16.8, 9.6, 5.3) km s $^{-1}$, which are close to the results of \citet{pi06}.
\begin{table}
\begin{minipage}{84mm}
\caption{Age-velocity dispersion relations derived from OCs in our present sample with errors in the spatial velocities less than 20 km s$^{-1}$.}\label{tb3}
\centering
\begin{tabular}{rcccc}
\hline
Age (Myr)&$\sigma_{U}$&$\sigma_{V}$&$\sigma_{W}$&$N$\\
\hline
  Age   $\le$  500           &  28.3 &  15.4 & 10.6 &  339\\
500 $<$ Age $\le$ 1000   &  29.1 &  13.8 &  11.9 & 25\\
1000 $<$ Age $\le$ 2000 &  31.7 &  23.4 &  17.7 & 16 \\
\hline
\end{tabular}
\end{minipage}
\end{table}

\section{Orbital parameters and their uncertainties}
\begin{table*}
\begin{minipage}{175mm}
\caption{The orbital parameters and their errors of 488 OCs in our sample calculated with AS91 model. The full version of this table is available in the online version of this article.}\label{tb4} \centering
\begin{tabular}{lccccccc}
\hline
name & $R_{a}$ & $R_{p}$ &$e$&$z_{\max}$& $T_{p}$&$T_{z}$&$J_{z}$\\
         & (kpc)&  (kpc)  &   & (kpc)    &     Myr&    Myr& (kpc km s$^{-1}$)\\
\hline Berkeley 59&$
8.6\pm0.4$&$8.5\pm0.3$&$0.01\pm0.03$&$0.09\pm0.02$&$239.3\pm
8.4$&$35.1\pm 1.6$&$-1883.3\pm 65.2$\\
Blanco 1   &$
8.2\pm0.1$&$7.5\pm0.2$&$0.05\pm0.02$&$0.26\pm0.05$&$221.1\pm
1.3$&$36.4\pm 1.3$&$-1726.2\pm 16.8$\\
Alessi 20  &$
8.3\pm0.1$&$7.2\pm0.1$&$0.07\pm0.02$&$0.03\pm0.01$&$217.0\pm
0.6$&$30.8\pm 0.3$&$-1694.8\pm  8.5$\\
ASCC 1
&$10.6\pm0.8$&$8.5\pm1.2$&$0.11\pm0.05$&$0.33\pm0.26$&$268.5\pm27.3$&$46.5\pm11.
9$&$-2059.9\pm209.0$\\
Mayer 1
&$10.0\pm0.9$&$8.5\pm0.3$&$0.08\pm0.03$&$0.50\pm0.24$&$261.8\pm16.5$&$50.0\pm10.
1$&$-2010.3\pm 94.4$\\
NGC 129    &$
8.9\pm0.2$&$7.4\pm0.4$&$0.10\pm0.02$&$0.28\pm0.12$&$229.3\pm
7.8$&$38.5\pm 4.5$&$-1772.2\pm 64.7$\\
ASCC 3     &$
9.0\pm0.2$&$7.8\pm0.4$&$0.07\pm0.02$&$0.21\pm0.04$&$235.6\pm
8.3$&$37.7\pm 2.5$&$-1836.6\pm 67.5$\\
NGC 225    &$
8.8\pm0.2$&$7.4\pm0.2$&$0.09\pm0.01$&$0.07\pm0.03$&$227.7\pm
4.5$&$33.3\pm 0.8$&$-1770.9\pm 35.5$\\
NGC 188    &$
9.3\pm0.6$&$8.1\pm0.8$&$0.07\pm0.04$&$0.79\pm0.21$&$246.7\pm17.3$&$57.1\pm
8.8$&$-1872.0\pm129.0$\\
IC 1590
&$10.0\pm0.6$&$9.2\pm0.6$&$0.04\pm0.02$&$0.33\pm0.17$&$268.7\pm17.0$&$46.5\pm
8.1$&$-2092.9\pm119.8$\\
\ldots&\ldots&\ldots&\ldots&\ldots&\ldots&\ldots&\ldots\\
\ldots&\ldots&\ldots&\ldots&\ldots&\ldots&\ldots&\ldots\\
\hline
\end{tabular}
\end{minipage}
\end{table*}
\subsection{The Galactic gravitational potential model}
In this study we employ the axisymmetric Galactic gravitational potential model of \citet[hereafter AS91]{as91}. This model consists of a spherical central bulge and a disc in the form proposed by \citet{mn75}, plus a massive, spherical halo extending to a radius of 100 kpc from the centre of the Galaxy. The total mass of the model is $9.0\times10^{11}\,M_{\sun}$ and the local total mass density at the solar position is $\rho_{0}=0.15\,M_{\sun}$ pc$^{-3}$. The rotation curve of this potential represents the current knowledge of galactic rotation in the Galaxy. This model is time-independent, completely analytical and very simple. The integration of the orbit, using this model, is very rapid and can achieve high numerical precision. The potential admits two conserved quantities, the total energy $E$ and the $z$-component $J_{z}$ of the angular momentum vector. This model has been used to derive the galactic orbits of OCs (C94), globular clusters \citep*[hereafter A06, A08]{od97,al06,al08}, and clump giants near the Sun (S08).

\subsection{The orbital parameters}
Using the data listed in Table~\ref{tb2}, the orbits are calculated backwards in time over a interval of 5 Gyr. Most clusters in our sample have ages less than 100 Myr, they do not even move one galactic orbit in the Galaxy. The integration time is chosen to ensure clusters can move more galactic orbits in the Galaxy and the averaged orbital parameters can be determined. For the integration we use the Bulirsch-Stoer algorithm of \citet{pr92}. The relative change in the total energy over the 5 Gyr integration time is of the order of $10^{-14}$ to $10^{-15}$.

The orbital parameters are listed in Table~\ref{tb4}. $R_{a}$ and $R_{p}$  are apogalactic and perigalactic distance from the Galactic centre, which are determined from the averaged maximum and minimum galactocentric distances of the cluster in the calculated Galactic orbit within the integration time of 5 Gyr. The orbital eccentricity $e$ is calculated as $e=(R_{a}-R_{p})/(R_{a}+R_{p})$, where $R_{a}$ and $R_{p}$ are averages. The maximum distance above the Galactic plane, $z_{\max}$, is also the averaged maximum vertical distances above the Galactic plane in the cluster's orbit within the given integration time. $T_{p}$ is the orbital period defined as the period of revolution around the $z$-axis. $T_{z}$ is the mean time interval of  the cluster to cross the Galactic plane from one $z_{\max}$ to the other one in the opposite direction.  Because of the right-handed orientation of the coordinate system adopted here, the negative $J_{z}$ of the $z$-component of the angular momentum vector corresponds to prograde rotation in the Galaxy and vice versa \citep{od97}.

\begin{figure*}
\begin{center}
\includegraphics[width=160mm]{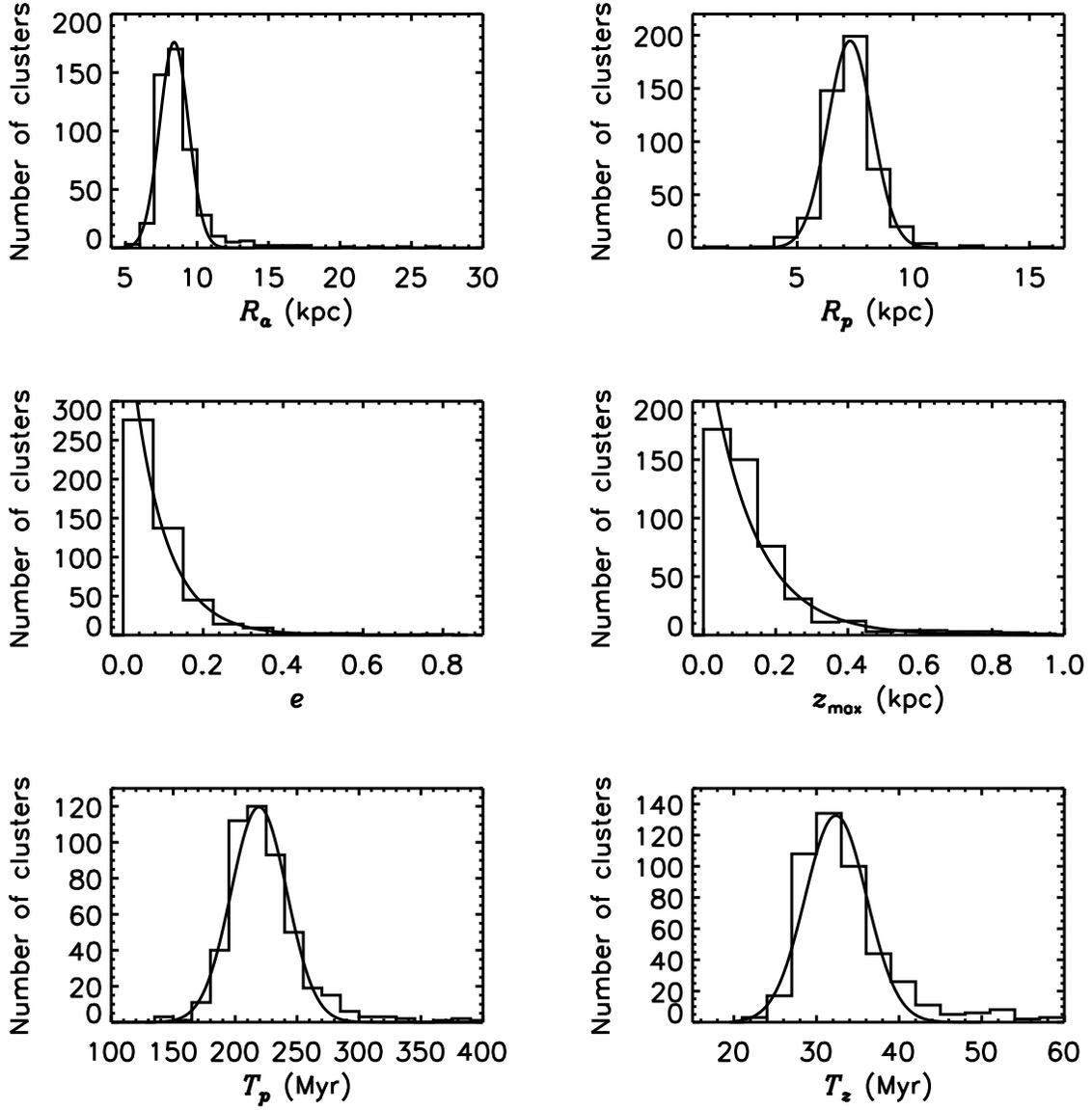}
\caption{ Distributions of derived orbital parameters calculated with AS91 model for OCs in our present sample.}\label{hobs}
\end{center}
\end{figure*}

In Fig.~\ref{hobs}, we plot the distributions of derived orbital parameters. In the panels of $z_{\max}$, $T_{p}$, and $T_{z}$, some very large values are not plotted. The distributions of $R_{a}$, $R_{p}$, $T_{p}$, and $T_{z}$ are fitted by Gaussian functions $\sim e^{-(x-\mu)^{2}/2\sigma^{2}}$, and the distributions of $e$ and $z_{max}$ are fitted by exponential functions $\sim e^{-x/\beta}$. The means and dispersions of the Gaussian functions for $R_{a}$, $R_{p}$, $T_{p}$, and $T_{z}$ are derived as: $\mu_{R_{a}}=8.40$ and $\sigma_{R_{a}}=1.00$ kpc, $\mu_{R_{p}}=7.28$ and $\sigma_{R_{p}}=0.96$ kpc, $\mu_{T_{p}}=219.3$ and $\sigma_{T_{p}}=22.5$ Myr, $\mu_{T_{z}}=32.3$ and $\sigma_{T_{z}}=3.7$ Myr, respectively. The parameters $\beta$ for the distributions of $e$ and $z_{max}$ are derived as: $\beta_{e}=0.08$ and $\beta_{z_{max}}=0.13$. The derived parameter $\beta$ and  Fig.~\ref{hobs} show that, for most of OCs, the orbital eccentricities $e$ are less than 0.1 and $z_{\max}$ less than 200 pc. In our sample, the minimum of the $R_{p}$ is bigger than 1 kpc and the mean of $R_{p}$ is $\sim 7.0$ kpc. Therefore,  the orbits of OCs in our present sample cannot be noticeably affected by a barred mass distribution in the Galactic centre within 1 kpc \citep[A06]{di99}. Fig.~\ref{hobs} also indicates that, in one orbital period, most clusters can cross the Galactic plane seven times.

\begin{figure*}
\begin{center}
\includegraphics[width=160mm]{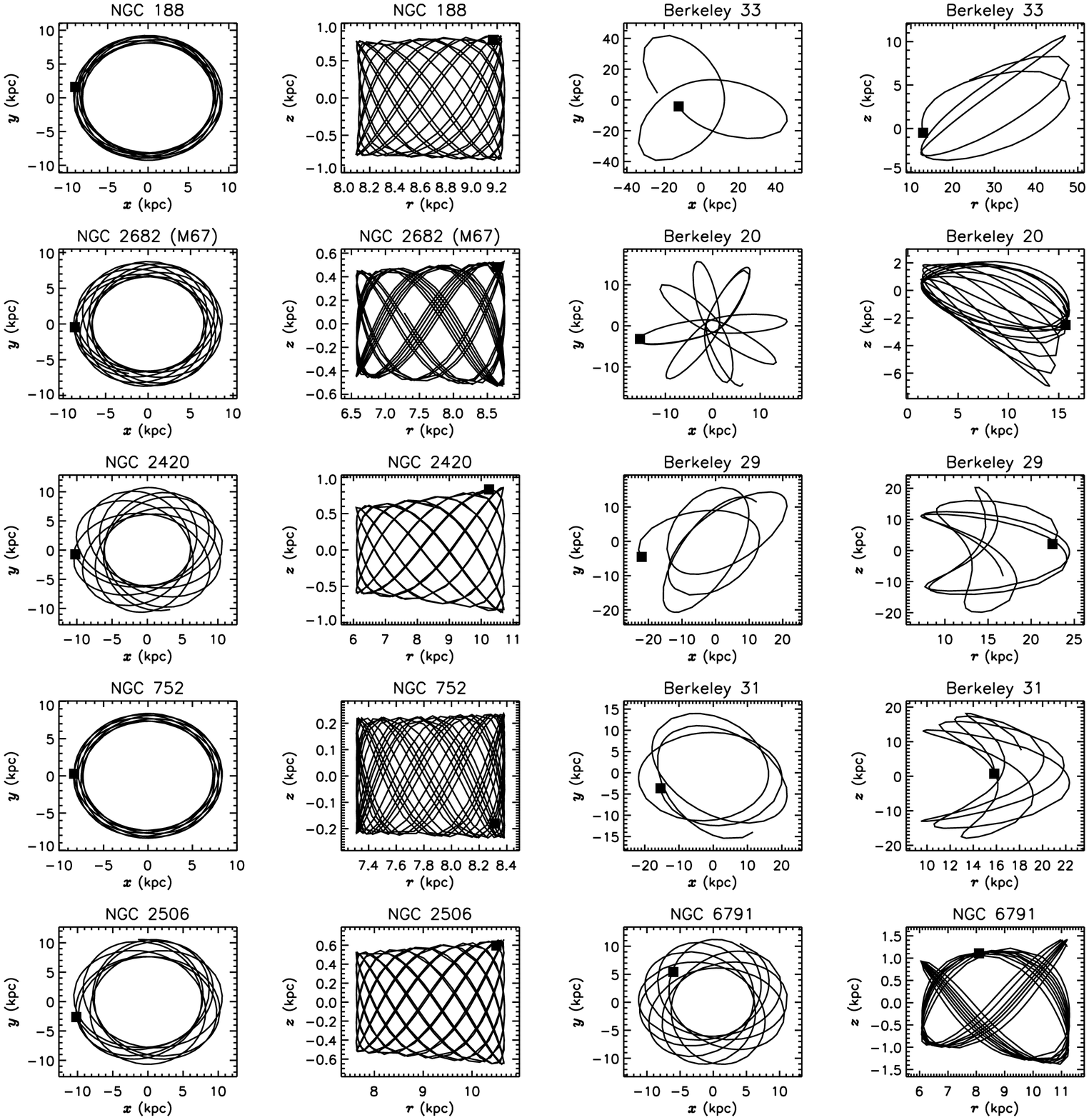}
\caption{Meridional Galactic orbits and orbits projected onto the Galactic plane in the time-interval of 2 Gyr for some OCs calculated with AS91 model. The filled square shows the present observed position for each cluster.}\label{obs91}
\end{center}
\end{figure*}

As a representative example, a few of orbits calculated with AS91 model in a time-interval of 2 Gyr are presented in Fig.~\ref{obs91}. For each cluster, the panel on the left shows the orbit projected onto the Galactic plane, while the panel on the right shows the meridional orbit. The filled square indicates the present observed position for each cluster. The orbits of NGC 188, NGC 2682, NGC 2420, NGC 752, and NGC 2506 are also calculated by C94 and \citet{fi95}. For other clusters presented in Fig.~\ref{obs91}, Berkeley 33 has the maximum of $R_{a}=48.3$ kpc, Berkeley 20 has the maximum eccentricity $e=0.81$,  Berkeley 29 and Berkeley 31 have the maximum values of $z_{\max}$, and NGC 6791 is among the most massive OCs known today \citep{fr99}.

All of the meridional orbits in Fig.~\ref{obs91} are of boxy-like type. Clusters move in the meridional plane within the limited areas almost filling the boxes symmetrically. But the meridional orbits of Berkeley 33 and  Berkeley 20 are not symmetric with respect to the Galactic plane. The orbits projected on the Galactic plane in Fig.~\ref{obs91} indicate the periodic motions of clusters more clearly.

\subsection{The errors in the orbital parameters}
The errors for the derived orbital parameters listed in Table~\ref{tb4} include two \textbf{types} of uncertainties affecting the derived results. The first one is the intrinsic variation of the orbital parameters within the 5 Gyr integration interval. The dispersions of the averaged orbital parameters are calculated over the number of galactic orbits, which indicate the intrinsic nature of the orbit, and this may be due to effects from chaos, and/or a complex distribution of orbit families \citep{di99}.

On the other hand, the main errors of the orbital parameters come from the observational uncertainties of the input data.  In the input data, a 20\% relative error is assumed for distance, the median of the relative errors in radial velocities is about 6\% and the median of the relative errors in proper motions is 23\%. The uncertainties in distance and proper motions are the main sources for the errors in the derived orbital parameters. The effects of observational uncertainties cannot be simply propagated into the derived orbital parameters \citep{od92}. Following \citet{di99}, the initial conditions are generated in a Monte Carlo fashion by adding Gaussian deviates to the observed absolute proper motions, radial velocities, and distances. The standard deviations are taken to be the errors for the input data listed in Table~\ref{tb1}. For each cluster, the errors for its orbital parameters have been calculated based on 1000 separate integrations.

Compared with the uncertainties in the derived orbital parameters due to the errors in the input data, the intrinsic uncertainties of the orbits within the given integration interval are very small. For each cluster, only considering the observational errors of the input data, the relative errors in the derived orbital parameters are also calculated. For each orbital parameter, the median of the relative errors for this parameter is calculated and listed as percentage in column 2 of Table~\ref{tb5}. Table~\ref{tb5} shows that the uncertainties of orbital eccentricity $e$ and the maximum distance from the Galactic plane $z_{\max}$ are much more affected by the observational errors. The large uncertainty of $e$ is propagated from the uncertainties of both $R_{a}$ and $R_{p}$. The reason for the large uncertainty of $z_{\max}$ is that, for most OCs, they move near the Galactic disc and have small vertical distances from the Galactic plane, so small changes in input data produce big relative changes of their orbits in the direction perpendicular to the disc.

\subsection{The effect of adopting different observational errors in input data on the derived orbital parameters}
Comparing the observational errors in the input data, the relative error in radial velocity is smaller than that in proper motion and distance. The errors in the derived orbital parameters are dominated by the observational errors in proper motions and distances. In order to estimate the effect of different observational accuracy in proper motions and distances on the derived orbital parameters, we repeated the Monte Carlo procedure mentioned above. We assumed 50\% and 80\% relative errors for the adopted proper motion and distance data, respectively. We also assumed  50\% and 80\% relative errors for both proper motion and distance all together.

There are six sets of simulations. In each set of simulation, except for the assumed different relative errors for proper motions or distances, the errors for other input data are the same as those considered in the previous section. For each set of simulations, we calculated the relative error for each orbital parameter, the medians of the relative errors are listed in Table~\ref{tb5a}. Table~\ref{tb5a} indicates that the relative errors in $e$ and $z_{\max}$ increase obviously with the increase of errors in input data. The uncertainties of $e$ and $z_{\max}$ are very sensitive to the errors of the input data just as Table~\ref{tb5} has indicated. Table~\ref{tb5a} also shows that, assuming the same relative error for proper motion and distance, the relative errors in the derived orbital parameters caused by the error of distance are bigger than those caused by the error of proper motion.

\section{Orbital parameters in different Galactic potential models}
\begin{table}
\begin{minipage}{84mm}
\caption{The relative error percentages of derived orbital parameters due to observational uncertainties and the relative difference percentages of derived orbital parameters by different Galactic potential models.}\label{tb5} \centering
\begin{tabular}{cccc}
\hline
Parameter& observational uncertainty&P90&FSC96\\
&(per cent)&(per cent)&(per cent)\\
\hline
$R_{a}$  & 2.9 & 0.0 &0.2  \\
$R_{p}$  & 3.5 &  0.5& 0.3\\
$e$          & 27.5 &  0.0&  0.0\\
$z_{\max}$&43.3&1.1& 21.2\\
$T_{p}$  &2.8&0.8&1.5\\
$T_{z}$   &3.9&2.9&31.5\\
\hline
\end{tabular}
\end{minipage}
\end{table}

\begin{table*}
\begin{minipage}{175mm}
 \caption{The relative error percentages of derived orbital parameters for adopted different observational uncertainties in proper motions $\mu$ and distances from the Sun $d_{\odot}$.}\label{tb5a} \centering
\begin{tabular}{lcccccc}
\hline
Parameter& \multicolumn{6}{c} {observational uncertainty}\\
&50\% in $\mu$&80\% in $\mu$& 50\% in $d_{\odot}$&80\% in
$d_{\odot}$&50\% in
$\mu$+$d_{\odot}$&80\% in $\mu$+$d_{\odot}$\\
$R_{a}$&3.9&4.8&6.8&11.5&7.5&14.5\\
$R_{p}$ &5.0&5.8&6.7&9.9&7.7&12.6\\
$e$&45.0&54.4&45.0&62.5&61.4&102.2\\
$z_{\max}$&76.8&112.9&81.5&130.0&121.0&271.7\\
$T_{p}$&3.7&4.3&5.8&9.7&6.5&12.3\\
$T_{z}$  &6.3&11.7&8.3&14.9&11.1&26.2\\
\hline
\end{tabular}
\end{minipage}
\end{table*}
We use two different potential models to calculate the differences in the derived orbital parameters. The first model is the one proposed by \citet[hereafter P90]{p90} and then used by \citet{di99} as their representative model to calculate the orbits of 38 globular clusters. This model consists of axisymmetric potential with three components: bulge, disc, and dark halo. The bulge is modeled as Plummer potential \citep{pl11}. The disc is the same as AS91 in the form of \citet{mn75} with different coefficients. The dark halo is modeled as logarithmic potential, which assures a flat rotation curve, but imply an infinite mass.

The second model is the one proposed by \citet*[hereafter FSC96]{fsc96} and used by N04 to calculate the orbits of nearby F and G dwarf stars. This axisymmetric potential also consists of three components: central core, disc, and dark halo. The potential of the central core is modeled by two spherical components, representing the bulge/stellar-halo and an inner core component. The disc is modeled using a combination of three analytical discs of \citet{mn75}. The potential of the dark halo is assumed to be logarithmic.

\begin{figure*}
\begin{center}
\includegraphics[width=160mm]{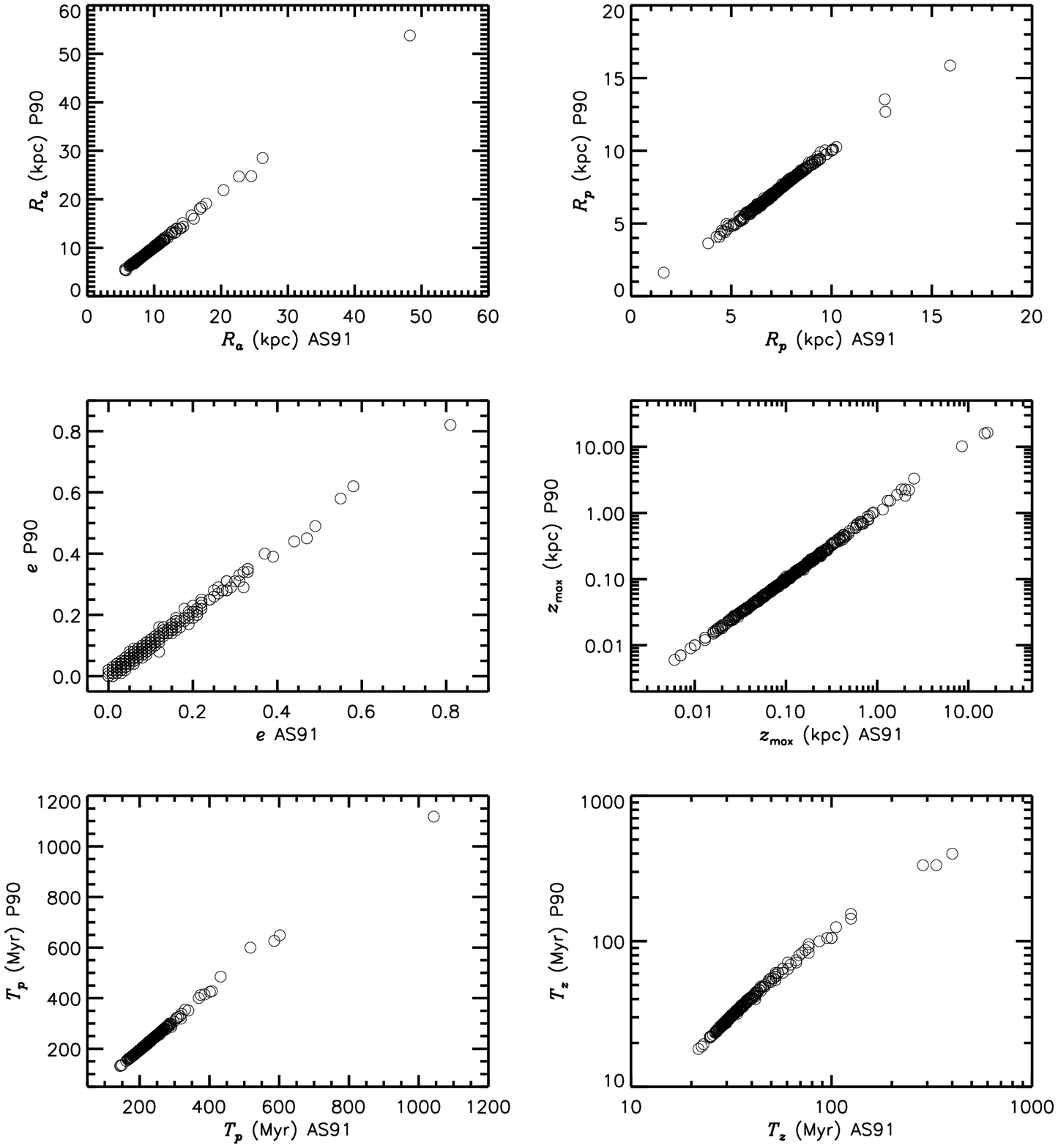}
\caption{Comparison of  derived orbital parameters with AS91 and P90 model.}\label{df90}
\end{center}
\end{figure*}

\begin{figure*}
\begin{center}
\includegraphics[width=160mm]{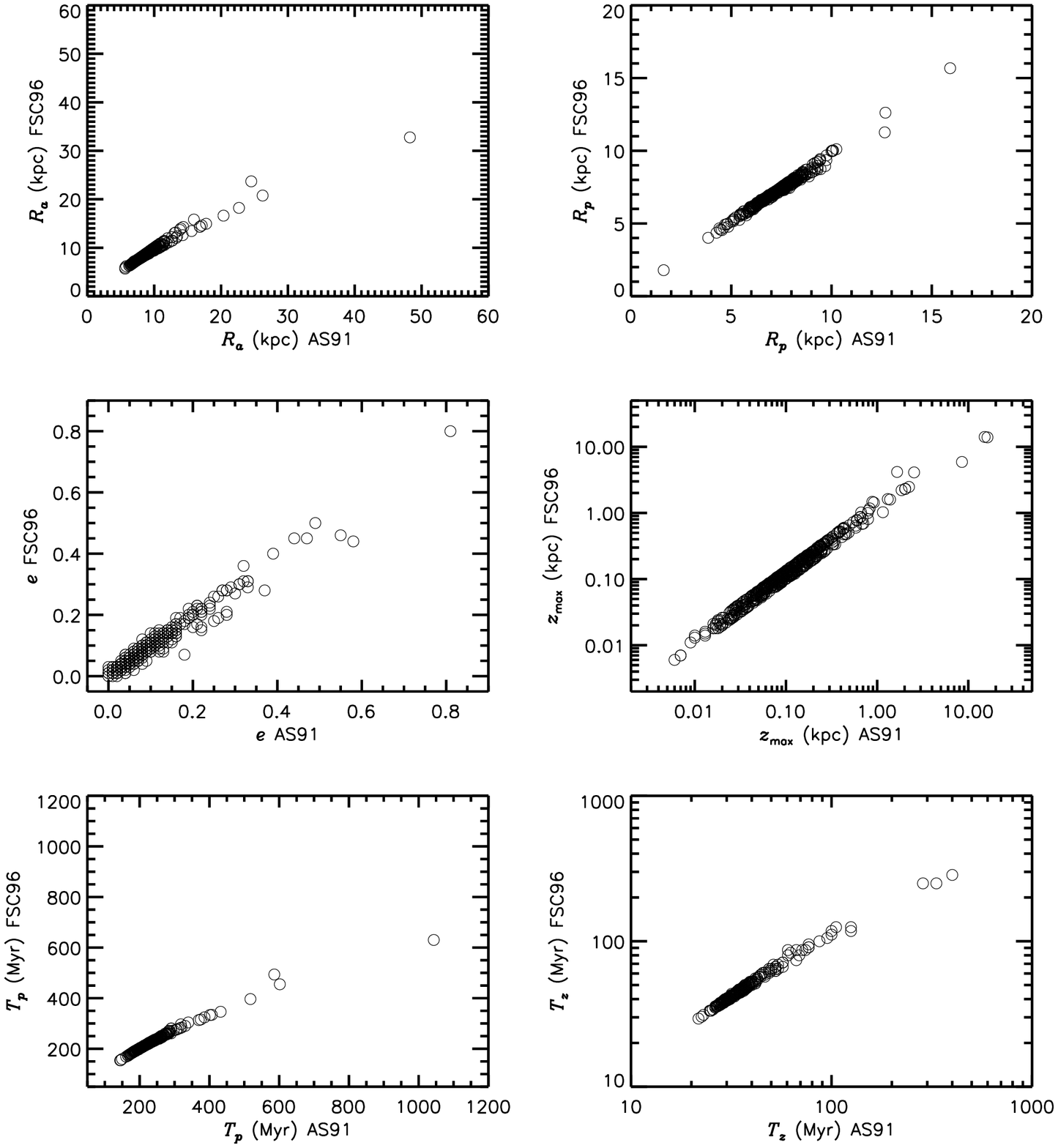}
\caption{ Comparison of derived orbital parameters with AS91 and FSC96 model.}\label{df96}
\end{center}
\end{figure*}

In Figs.~\ref{df90} and~\ref{df96}, we compare the orbital parameters derived from P90 and FSC96 models with those derived from AS91 model. Fig.~\ref{df90} shows that the orbital parameters derived from P90 model are very consistent with those derived from AS91 model, and no systematic differences can be found. For FSC96 model, Fig.~\ref{df96} shows that $R_{p}$ and $e$ are consistent between FSC96 and AS91 models, and there are no systematic differences in these two parameters. But for $R_{a}> 15 $ kpc,  the derived $R_{a}$  and $T_{p}$ for FSC96 model are systematically smaller than those derived from AS91 model.  Fig.~\ref{df96} also indicates that the derived $z_{\max}$ and $T_{z}$ for FSC96 model are systematically bigger than those derived from AS91 model.

\begin{figure*}
\begin{center}
\includegraphics[width=160mm]{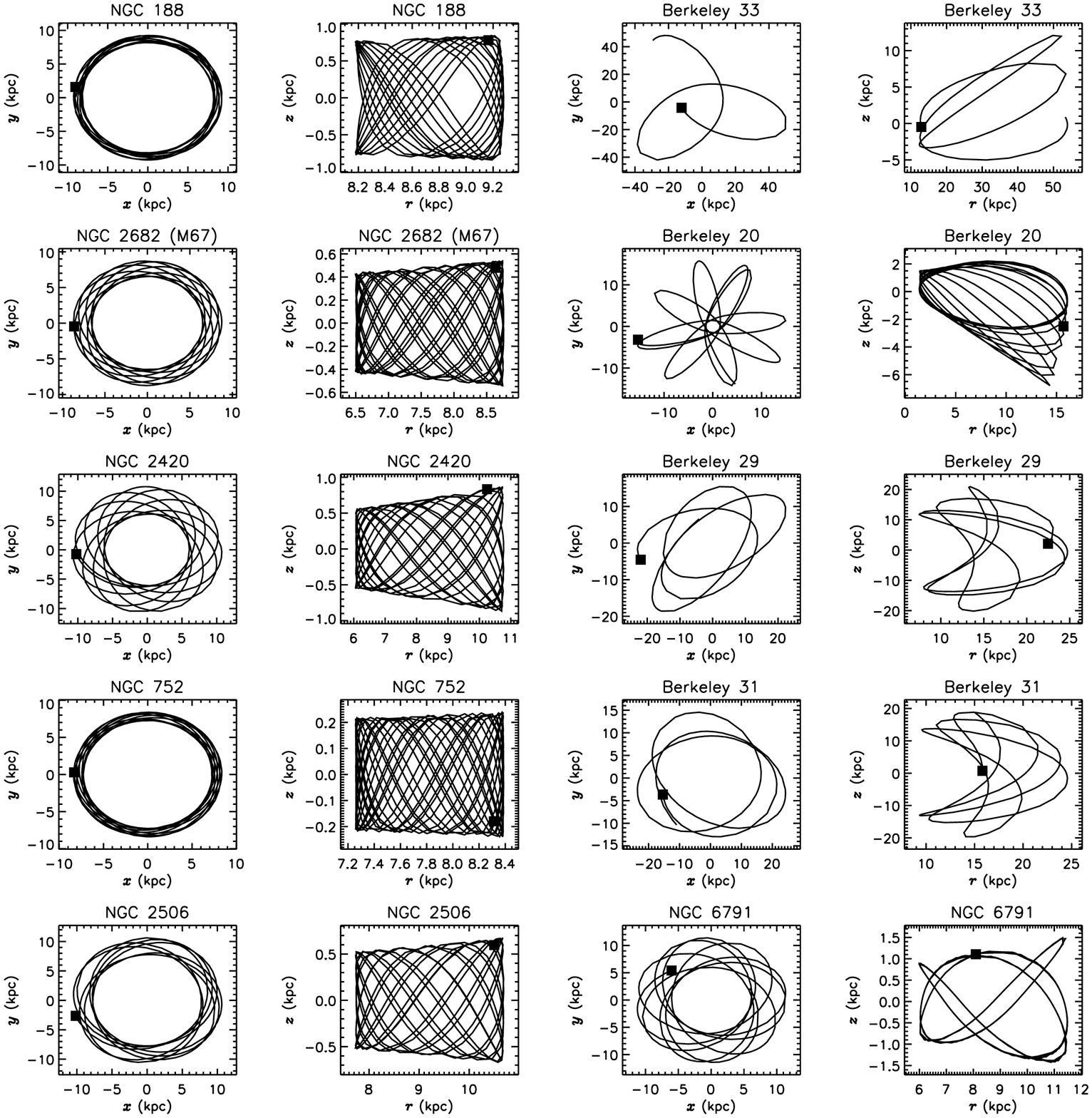}
\caption{Meridional Galactic orbits and orbits projected onto the Galactic plane in the time-interval of 2 Gyr for some OCs calculated with P90 model. The filled square shows the present observed position for each cluster.}\label{obs90}
\end{center}
\end{figure*}

\begin{figure*}
\begin{center}
\includegraphics[width=160mm]{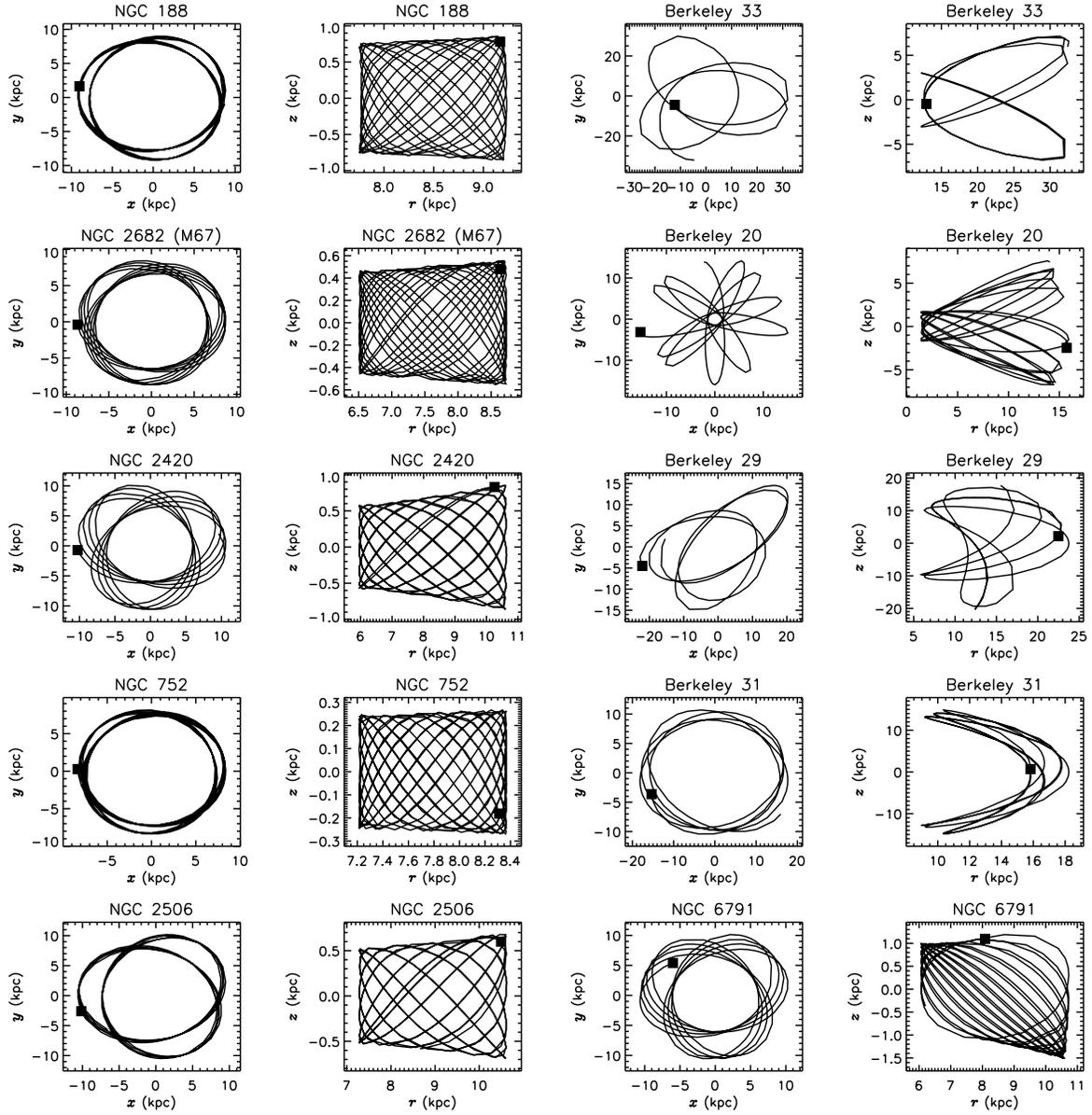}
\caption{Meridional Galactic orbits and orbits projected onto the Galactic plane in the time-interval of five times orbital periods for some OCs calculated with FSC96 model. The filled square shows the present observed position for each cluster.}\label{obs96}
\end{center}
\end{figure*}

In Figs.~\ref{obs90} and ~\ref{obs96}, for P90 and FSC96 models, we show the orbits for the same clusters as presented in Fig.~\ref{obs91}. Fig.~\ref{obs90} indicates that the orbital shapes for most of clusters, including the asymmetric meridional orbits of Berkeley 33 and  Berkeley 20, are very similar between P90 and AS91 models. Fig.~\ref{obs96} shows that for most of clusters, the orbital shapes derived from FSC96 model are different from those derived from AS91 model. The meridional orbits of Berkeley 33 and Berkeley 20 are symmetric in FSC96 model.

For each cluster, the relative difference in each orbital parameter from different models is calculated as $(P_{\textrm{(P90 or FSC96)}} - P_{\textrm{AS91}})/P_{\textrm{AS91}}$, where $P_{\textrm{P90}}$, $P_{\textrm{AS91}}$, and $P_{\textrm{FSC96}}$ represent the derived orbital parameters by P90, AS91, and FSC96 models respectively. For each orbital parameter, the median of the relative difference between the given models and AS91 model is listed in Table~\ref{tb5}. Table~\ref{tb5} shows that the relative differences of the derived orbital parameters for P90 model are smaller than those for FSC96 model. Especially for $z_{\max}$ and $T_{z}$, the results derived from FSC96 model are very different from those derived from AS91 and P90 models. The consistent orbital parameters derived from AS91 and P90 models are due to the very similar mass distributions of these two models. The main difference between AS91 and P90 models is the bulge model, whereas bulge model dominates the orbit of a cluster only within 1 kpc from the Galactic centre. No clusters in our present sample can move to this range. The FSC96 model is very different from AS91, the total mass interior to the same  galactocentric distance $R$ in the FSC96 model is much larger than that in the AS91 model when $R > 10 $ kpc. The larger mass of FSC96 model can control the cluster to move within smaller apogalactic distance. On the other hand, the mass of disc in FSC96 model is smaller than that in AS91 model, clusters can move to more larger vertical distances in FSC96 model.

Table~\ref{tb5} also shows that, for most of derived orbital parameters, the relative differences due to different potential models are smaller than those from observational errors. But for FSC96 model, the relative differences in $T_{z}$ are much larger than the uncertainties due to observational errors of input data, which indicates the major effect of the disc model of FSC96.
\section{Characteristics of the orbital parameters}

\subsection{Relations between $R_{a}$ --- $R_{\textrm{GC}}$, $R_{p}$ --- $R_{\textrm{GC}}$,  and $z_{\max}$ --- $|z|$.}
\begin{figure*}
\begin{center}
\includegraphics[width=160mm]{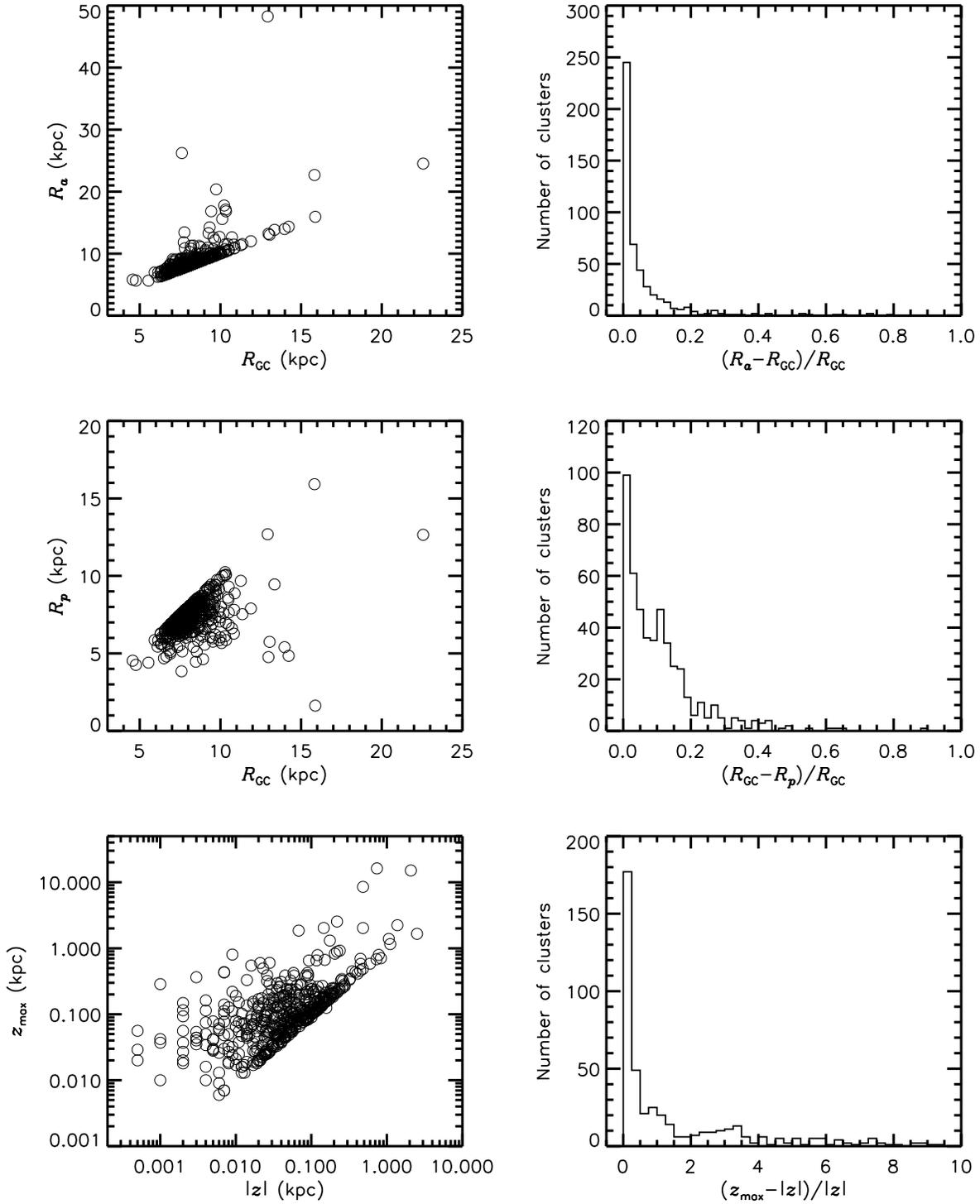}
\caption{Relations between the orbital parameters $R_{a}$, $R_{p}$, $z_{\max}$ of OCs in our present sample and their observed data $R_{GC}$, $|z|$.}\label{rzs}
\end{center}
\end{figure*}

The left panels of Fig.~\ref{rzs} show the $R_{a}$ vs $R_{\textrm{GC}}$, $R_{p}$ vs $R_{\textrm{GC}}$, and $z_{\max}$ vs $|z|$ diagrams, where $|z|$ is the current observed vertical distance from the Galactic disc. The panel of $R_{a}$ vs $R_{\textrm{GC}}$ diagram indicates that, for most of clusters in present sample, their current observed positions are very close to their apogalacticons. Whereas,  the panel of $R_{p}$ vs $R_{\textrm{GC}}$ diagram indicates that most of clusters are far away from their perigalacticons. The panel of $z_{\max}$ vs $|z|$ diagrams indicates that most of clusters do not arrive at their maximum distances from the disc and are crossing the Galactic plane.

In the right panels of Fig.~\ref{rzs}, we plot the histograms of the relative differences for corresponding parameters in the left panels. The top panel shows the relative difference between $R_{a}$ and $R_{\textrm{GC}}$ and indicates that, at present, about 70\% clusters are moving within only about 5\% distances from their apogalacticons. The median of the relative differences between $R_{a}$ and $R_{\textrm{GC}}$ is 2.0\%. The middle panel shows the relative difference between $R_{p}$ and $R_{\textrm{GC}}$, and indicates a significantly larger relative difference with respect to that between $R_{a}$ and $R_{\textrm{GC}}$. The median of the relative differences between $R_{p}$ and $R_{\textrm{GC}}$ is 8.1\% which is larger than that between $R_{a}$ and $R_{\textrm{GC}}$. The bottom panel shows the relative differences between $z_{\max}$ and $|z|$, and indicates very large differences between these two parameters. The median of the relative differences is 68.8\%. The mean time $T_{z}$ for our present sample is very short, so the time for a cluster moving near its $z_{max}$ is short, which is the possible reason for the large differences between the observed $z$ and the orbital parameter $z_{max}$.

\begin{figure*}
\begin{center}
\includegraphics[width=140mm]{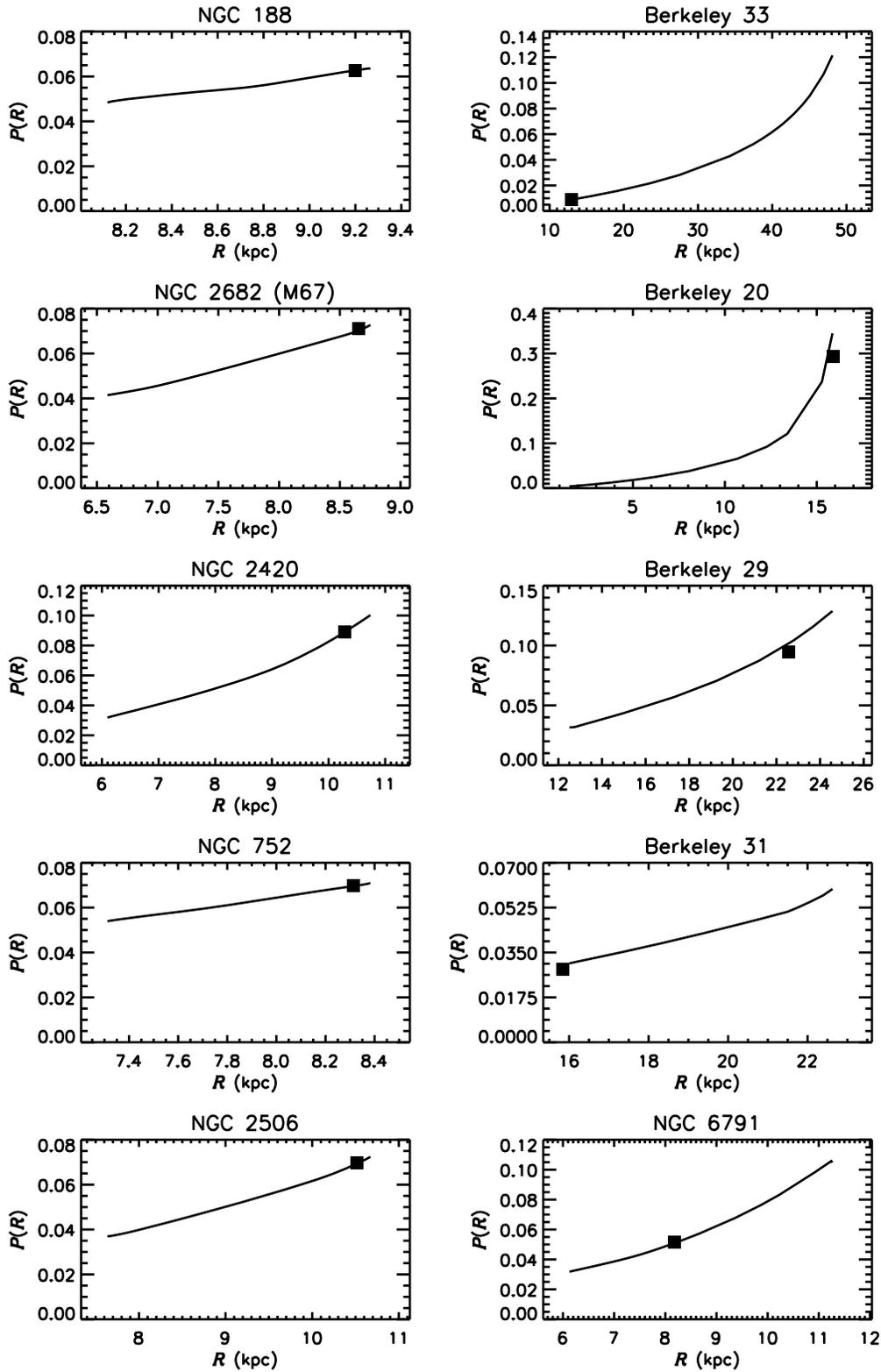}
\caption{The detection probability distributions $P(R)$ for some OCs in our present sample.The filled square is the present observed position for each cluster.}\label{prs}
\end{center}
\end{figure*}

For two clusters NGC 2682 and NGC 2420 whose orbital eccentricities are larger than 0.1, C94 calculated the probability of finding the clusters at the given galactocentric distances and found that the detection probability is the largest at the cluster's apogalacticon.  Following C94, we also define a probability function $P(R) \propto \frac{1}{T_{p}}\frac{R}{\nu(R)}$ to calculate the probability of a cluster to be observed at the galactocentric position $R$ during the orbital period $T_{p}$, where $\nu(R)$ is the cluster's velocity at $R$. Fig.~\ref{prs} shows the probability distributions $P(R)$ for the same clusters as in Fig.~\ref{obs91}. The filled square in Fig.~\ref{prs} indicates the present observed position for each cluster. Fig.~\ref{prs} shows that the detection probability $P(R)$ increases with the galactocentric distance $R$, the largest detection probability is at the cluster's apogalacticon. The derived probabilities for other clusters also indicate that the detection probability for a cluster at its apogalacticon is the largest one during its orbital period, which is consistent with the result of C94. Because the detection probability at the apogalacticon for a cluster is the largest during its orbital period,  it is easier to find a cluster near its apogalacticon just as Fig.~\ref{rzs} indicates. 

\subsection{The radial metallicity gradient}
The OCs have long been used as tracers of radial metallicity gradients in the Galactic disc. Since the early work by \citet{j79}, others have found general agreement in the existence and magnitude of the trend. Most investigators have found gradients of $-0.06$ to $-0.09$ dex kpc$^{-1}$ over a range of distances from 7 to 16 kpc from the Galactic centre \citep*{tw97, fr99}. Recently, \citet{fr02} presented metallicities for a sample of 39 intermediate age and old OCs based on an updated abundance calibration of spectroscopic indices. They found a metallicity gradient of $-0.06\pm0.01$ dex kpc$^{-1}$ over a range of galactocentric distances of 7 to 16 kpc. \citet{ch03} compiled a OCs catalogue of 119 objects with ages, distances, and metallicities available,  which led to a metallicity gradient of $-0.063\pm0.008$ dex kpc$^{-1}$, similar to the result derived by \citet{fr02} from a homogeneous sample.

\begin{figure*}
\begin{center}
\includegraphics[width=160mm]{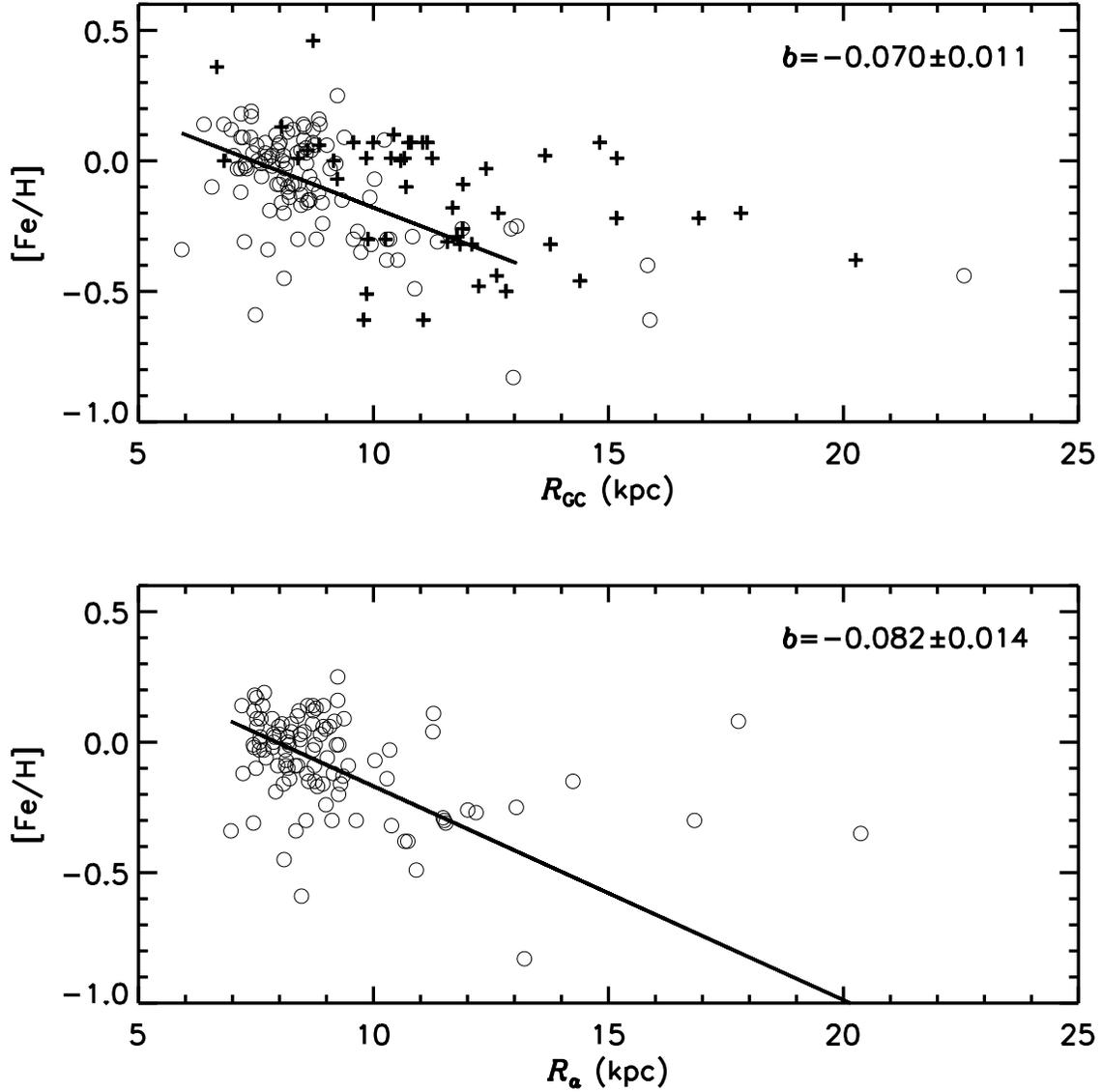}
\caption{Metallicity gradients for OCs. Clusters included in our present sample are plotted as open circles, clusters not included in our present sample but with [Fe/H] data are plotted as plus signs. Top panel: The metallicity gradient of OCs is derived based on the currently observed galactocentric distances of OCs $R_{\textrm{GC}}$; bottom panel: The metallicity gradient of OCs is derived based on apogalactic distances $R_{a}$. The straight-lines are the best-fitting results of the metallicity gradients for OCs in our present sample with $R_{\textrm{GC}} < 13.5$ kpc and the slopes $b$ for each line, i.e.\, the gradients, are labeled in each panel.}\label{fehr}
\end{center}
\end{figure*}

The [Fe/H] vs current observed galactocentric distances $R_{\textrm{GC}}$ diagram is plotted in the top panel of Fig.~\ref{fehr}. 109 clusters with [Fe/H] data listed in our present sample are plotted as open circles, 48 clusters not listed in our present sample but with [Fe/H] data are plotted as plus signs. There are 12 clusters in the range of $R_{\textrm{GC}} > 13.5$ kpc, only 3 clusters are included in our present sample. Considering the observational errors in [Fe/H] and $R_{\textrm{GC}}$, we perform a linear least-square fitting to the clusters listed in our sample with  $R_{\textrm{GC}} < 13.5$ kpc. For each cluster, we assign the typical observational uncertainty of 0.15 dex to their [Fe/H] data. The errors of $R_{\textrm{GC}}$ are calculated from the data listed in Table~\ref{tb2}. We get a gradient of $-0.070\pm0.011$ dex kpc$^{-1}$. We also perform the same fitting to all clusters in the range of $R_{\textrm{GC}} <13.5$ kpc, we get  a gradient of $-0.069\pm0.008$ dex kpc$^{-1}$. The fitted straight-line for the clusters listed in our present sample within $R_{\textrm{GC}} < 13.5$ kpc is plotted in the same panel of Fig.~\ref{fehr} and the slope $b$ of this line is also labeled.

In the \textbf{most} recent study, \citet{ch07} derived a radial metallicity gradient of $-0.058\pm0.006$ dex kpc$^{-1}$ based on a sample of 144 OCs. Our derived radial metallicity gradient of $-0.07$ is slightly smaller than the previous results \citep{fr02, ch03, ch07}. Based on 45 OCs with high-resolution spectroscopy, \citet{ma09} found a steep metallicity gradient for clusters with $R_{\textrm{GC}} < 12.0$ kpc and a plateau for clusters at larger galactocentric distances. Fig.~\ref{fehr} indicates that the metallicity distribution of clusters with $R_{\textrm{GC}} > 13.5$ kpc is flat and there is no significant radial metallicity gradient within this distance range, which is consistent with the result of \citet{ma09}. In our present sample, only three OCs have galactocentric distances greater than 13.5 kpc (the open circles in Fig.~\ref{fehr}). In order to diminish the effect of the small sample at large galactocentric distances, we only consider OCs with $R_{\textrm{GC}} < 13.5$ kpc in our present sample to derive the radial metallicity gradient. The maximum galactocentric distances of the OCs in the previous studies extend to 17 kpc and some clusters within the flat metallicity distribution were included in their samples\citep{fr02, ch03, ch07}, which is the main reason \textbf{why} they obtained \textbf{larger} radial metallicity gradient. If we use all clusters with $R_{\textrm{GC}} < 17.0$ kpc, we get a gradient of $-0.056\pm0.007$ dex kpc$^{-1}$ which is consistent with the \textbf{most} recent result of \citet{ch07}.

The radial metallicity gradient of the OCs provides strong constraints on the formation and evolution of the Galaxy. Detailed models of Galactic chemical evolution have been improved over \textbf{the} last decades and most models can reproduce the presently observed radial metallicity distribution of the Galaxy\citep[and references therein]{ma09}. More recently, \citet{fu09} considered various mechanisms including infall, star formation and delayed disk formation to find the effect of each mechanism on their derived galactic chemical evolution model. They found that using the star formation rate(SFR) of the modified Kennicutt law, their model can properly predict both the current metallicity gradient and its time evolution. But, their best model also predicts that the outer disc has a steeper gradient than the inner disc, which is contrary to the result derived from OCs. \citet{ma09} adopted an inside-out formation model of the Galactic disc to reproduce the radial metallicity gradient of the OCs. In their model, the infall of gas is represented by an exponential law combined with the distribution of gas in the halo. The inner regions are rapidly evolving due to the higher infall and SFR, while the outer parts evolve more slowly. Their model can reproduce the main features of the metallicity gradient and the evolution of the OCs. In order to better reproduce the metallicity plateau at large galactocentric distances, an additional uniform inflow per unit disc area should be considered, but it is difficult to reconcile with the present-day radial distribution of the SFR. \textbf{A sequence of merging episodes} in the past history of the Galaxy would explain the outer metallicity plateau of the OCs\citep{ma09}.

In our previous discussion, we have showed that the apogalacticon is the place where a cluster spends the largest fraction of its life. In the bottom panel of Fig.~\ref{fehr}, we plot the [Fe/H] vs $R_{a}$ diagram for clusters listed in our sample within $R_{\textrm{GC}} < 13.5$ kpc. We also perform a linear least-square fitting to the data considering the observational errors in [Fe/H] and $R_{a}$. The errors of $R_{a}$ are taken from Table~\ref{tb3}. We plot the fitted straight-line in the same panel and also label the slope $b=-0.082\pm0.014$ dex kpc$^{-1}$ in the panel. This result indicates that, for clusters within $R_{\textrm{GC}} < 13.5$ kpc, the observed metallicity gradient at present is similar to that derived from the most probable observed positions of the clusters, which is consistent with the result of C94.

\subsection{Comparison of orbital eccentricities for different populations}
\begin{figure*}
\begin{center}
\includegraphics[width=160mm]{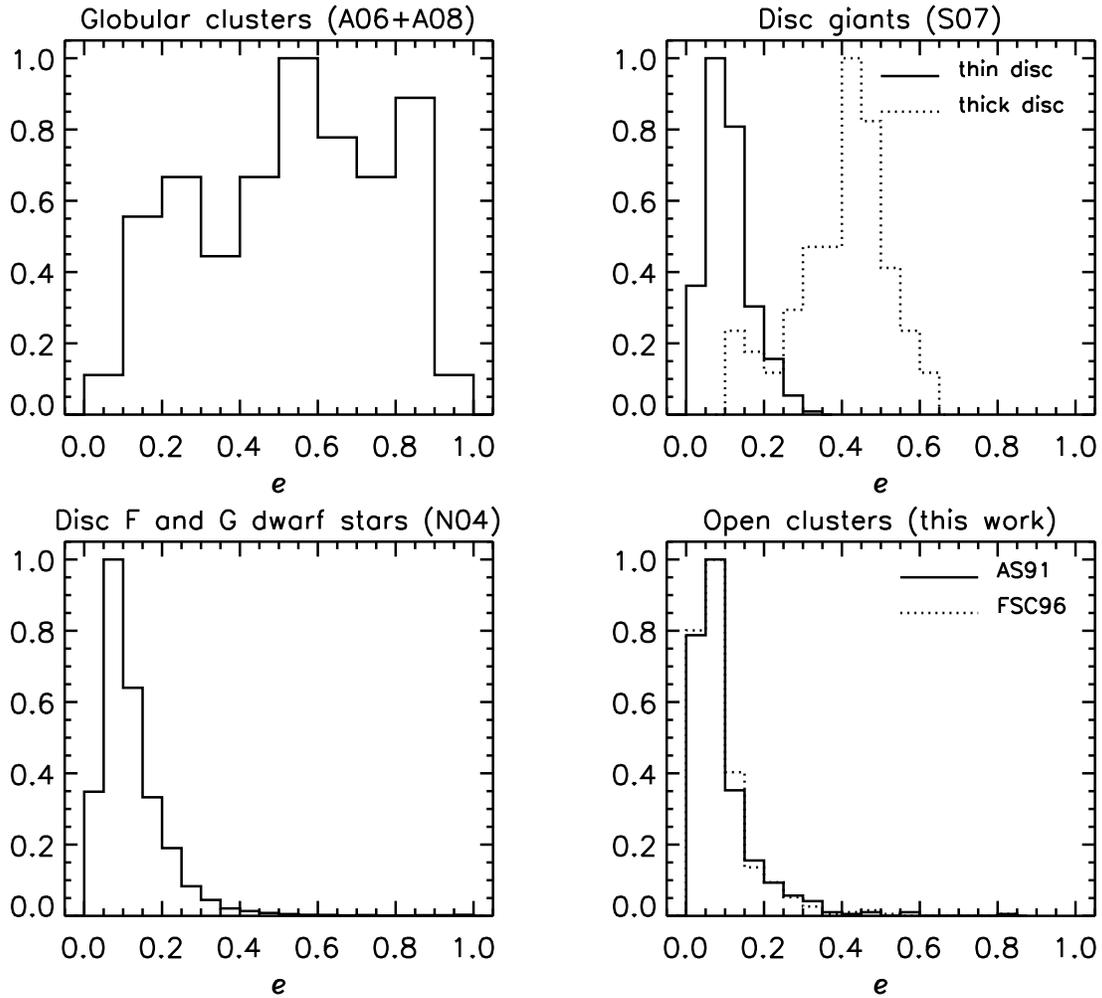}
\caption{Comparison of  orbital eccentricities $e$ for different populations: globular clusters (A06+A08), disc giants (S08), disc F and G dwarf stars (N04), and OCs (this work).}\label{hcom}
\end{center}
\end{figure*}

Fig.~\ref{hcom} shows the histograms of orbital eccentricities for different populations: globular clusters (top left panel), disc giants (top right panel), disc F and G dwarf stars (lower left panel), and OCs (lower right panel). \textbf{Each histogram is} normalized to its maximum value in the distribution. The orbital parameters for 54 globular clusters calculated with AS91 model are taken from A06 and A08. The orbital parameters of disc clump giants calculated with AS91 model are taken from S08. The orbital parameters of disc F and G dwarf stars calculated with FSC96 model are taken from N04.

Globular clusters represent the halo population which is primarily a system with a large velocity dispersion, and a wide range of orbit characteristics is expected \citep{di99}. This is indeed what the top left panel of Fig.~\ref{hcom} shows. There is a large range in orbital eccentricities of globular clusters, which can get as low as 0.1, but, in the mean, the eccentricities are high with an average of $\sim 0.5$.

S08 assigns to each star its probability to belong to the thin disc, the thick disc, the Hercules stream and the halo based on the basis of its ($U$, $V$, $W$) velocity and their velocity ellipsoids. In the top right panel of Fig.~\ref{hcom}, we show the histograms of orbital eccentricities for stars with probabilities higher than 80\% to belong to the thin disc (solid lines) and stars with probabilities higher than 80\% to belong to the thick disc (dot lines). Fig.~\ref{hcom} indicates that no thin disc giant has eccentricity greater than 0.3 and no thick disc giant has eccentricity less than 0.1. The orbital eccentricities of the thin and the thick disc giants are overlapped between 0.1 and 0.3. The average of orbital eccentricities for the thin disc giants is $\sim0.1$, and the one for the thick disc giants is $\sim0.4$, which is less than the average of orbital eccentricities for globular clusters.

The histogram of orbital eccentricities for disc F and G dwarf stars in the lower left panel of Fig.~\ref{hcom} shows that most of stars belong to the thin disc whereas a fraction of stars belong to the thick disc. If we assume that stars with eccentricities greater than 0.3 belong to the thick disc, the thick-disc fraction is about 3.6\% which is close to the value of 2.9\% derived by \citet{ho07} from the same sample with different method.

The lower right panel of Fig.~\ref{hcom} shows the histograms of orbital eccentricities for OCs in our present sample calculated with AS91 model (solid lines) and FSC96 model (dot lines). The distributions of eccentricities in these two models are very similar. The distribution of orbital eccentricities for OCs is similar to that of disc F and G dwarf stars which indicates that most of clusters belong to the thin disc and a fraction of thick-disc clusters exist in our sample. We assume the same limited value of 0.3 for orbital eccentricities to distinguish thin-disc and thick-disc clusters, and find 3.7\% clusters in our sample are probably thick-disc clusters.

In our present sample, the $z_{\max}$ values of Berkeley 29, Berkeley 31, and Berkeley 33 are $15.1\pm15.4$, $16.2\pm 12.2$, and $8.5\pm 9.3$ kpc, which are much larger than those of the other clusters in the sample. The maximum of $z_{\max}$ of the other clusters is less than 3 kpc and the observed maximum of $z$ is 2.1 kpc for Berkeley 29. The very large errors of $z_{\max}$ for these three clusters indicate that their orbits are very uncertain and more observations are needed to improve the precision of the input data for these clusters. Berkeley 20 has the maximum of $e=0.81\pm0.17$ in our present sample which indicates it is a halo cluster. But this cluster has an age of 6.0 Gyr and [Fe/H]$=-0.61\pm0.14$ indicating it is a thick disc cluster. If we adopt the limited value $e=0.3$ with $3\sigma$ from the mean, the cluster can be identified as a member of the thick disc population only based on orbital eccentricity $e$. More observations are also needed to improve the orbital parameter for this cluster.

\subsection{Comparison with previous results and the open cluster complexes}
\begin{figure*}
\begin{center}
\includegraphics[width=160mm]{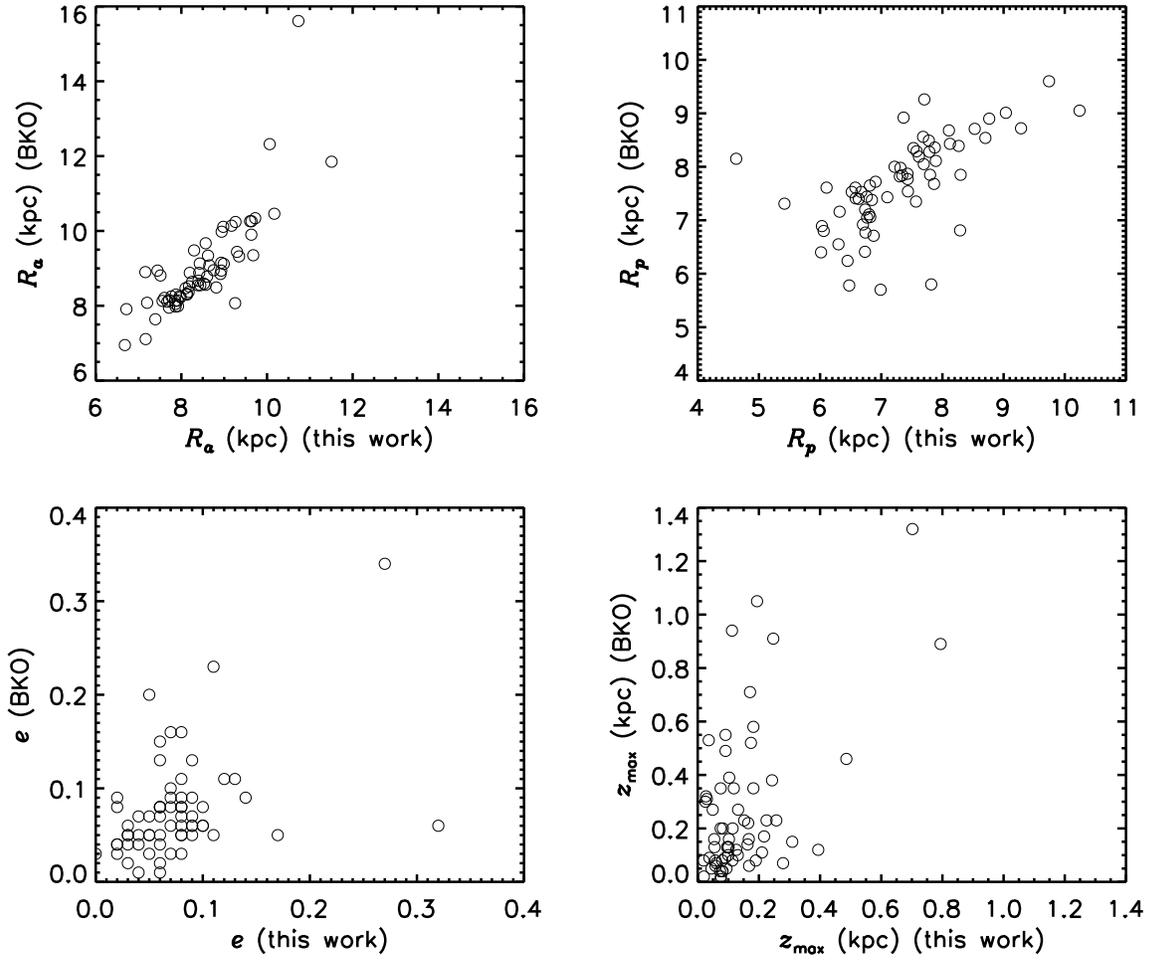}
\caption{The derived orbital parameters $R_{a}$, $R_{p}$, $e$, and $z_{\max}$ with AS91 model in this work are compared with those derived by BKO.}\label{comp}
\end{center}
\end{figure*}

There are two studies which calculated the orbits for a large samples of OCs and the orbital parameters were also derived.  BKO calculated the orbits for 69 OCs and the orbital parameters $R_{a}$, $R_{p}$, $e$, and $z_{\max}$ were also derived for those clusters. In Fig.~\ref{comp}, for common clusters, we compare the derived orbital parameters in AS91 model with those derived by BKO. Fig.~\ref{comp} indicates that, for most clusters, the derived $R_{a}$, $R_{p}$, and $e$ are consistent with those derived by KBO. The maximum difference in $R_{a}$ is for NGC 2420. The radial velocity for this cluster is $73.6$ km s$^{-1}$ in our present study, but the value of  $115$ km s$^{-1}$ was adopted by BKO. The large difference in radial velocity make the large difference in $R_{a}$ for NGC 2420. The maximum difference in $R_{p}$ and $e$ is for NGC 7789. The radial velocity of $-64$ km s$^{-1}$ is adopted for this cluster in our study, BKO adopted the value of $-32$ km s$^{-1}$.  The large difference in $R_{p}$ and $e$  for NGC 7789 is also due to the large difference in radial velocity data. For most of OCs, the orbital parameter $z_{max}$ in our present study are less than those derived by BKO. The systematic difference in $z_{max}$ is due to the different disc models adopted by this study and by BKO. 

\citet{pi06} calculated the orbits for a sample of 148 OCs within $d_{xy} \le 0.85$ kpc and the mean orbital parameters are listed in their Table 2: $\mu_{R_{a}}=8.631$ kpc, $\mu_{R_{p}}=6.706$ kpc, $\mu_{e}=0.127$, and $\mu_{z_{max}}=0.260$ kpc. For the clusters in our present sample within the same distance range, the corresponding mean orbital parameters calculated with AS91 model are derived as: $\mu_{R_{a}}=8.289$ kpc, $\mu_{R_{p}}=7.378$ kpc, $\mu_{e}=0.059$, and $\mu_{z_{max}}=0.084$ kpc. Because the input data for most of clusters in this study are similar to those in the study of \citet{pi06}, the differences of the derived orbital parameters are mainly due to the adopted different Galactic models. \citet{pi06} used the Galactic model of \citet{sa79}, the analytic solution of the orbital parameter can be derived from their model. In comparison, the much larger disc mass in AS91 model is responsible for the large difference in the derived value of $z_{max}$ with respect to the other model.

Many surveys have found that the structure of the Galaxy is more complex than previously thought, and about 10 moving groups can be identified from nearby stars with heliocentric distances less than 100 pc \citep{bo09}. Using OCs with the projected distance onto the Galactic plane $d_{xy} \le 0.85$ kpc from the Sun, and based on the surface density distribution and the tangential velocity distribution of those clusters, \citet{pi06} identified four open cluster complexes (OCCs). The numbers of kinematic member clusters for the four OCCs: OCC 1, OCC 2, OCC 3, and OCC 4 are 23, 27, 8, and 9 \citep{pi06}. In our present sample, for the corresponding OCCs, the numbers of member clusters identified by \citet{pi06} whose orbits can be determined are 20, 13, 3, and 4.

\begin{figure*}
\begin{center}
\includegraphics[width=160mm]{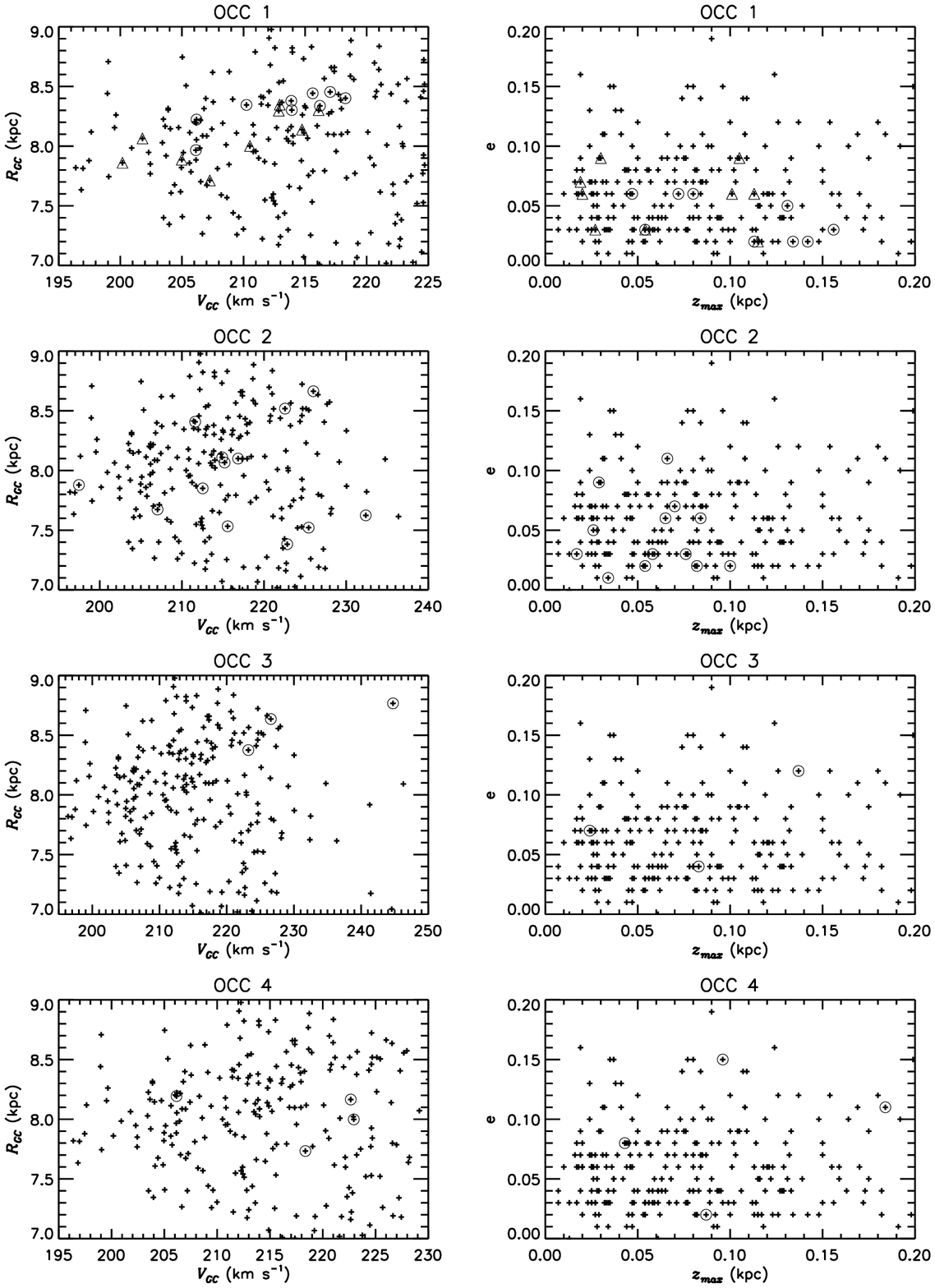}
\caption{The orbital parameter distributions of four open cluster complexes (OCCs) identified by \citet{pi06}. Open circles are member clusters for each OCC and crosses are clusters with heliocentric distance less than 1.3 kpc. In the panels of OCC 1, open circles represent member clusters \textbf{younger} than 30 Myr and triangles represent member clusters with age between 30 and 80 Myr.}\label{occs}
\end{center}
\end{figure*}

Those OCCs were detected from their overdensity in the spatial distribution of OCs in the Solar neighbourhood and the membership of the member clusters were determined only based on their tangential velocities\citep{pi06}. In Fig.~\ref{occs}, we use the $V_{GC}$ --- $R_{GC}$ and $z_{max}$ --- $e$ diagrams to \textbf{attempt to recover} those OCCs, where $V_{GC}$ is the total galactocentric velocity of cluster. The crosses in Fig.~\ref{occs} represent clusters with heliocentric distance less than 1.3 kpc, the open circles represent member clusters in each OCC. In the panels of OCC 1, the open circles represent member clusters with age less than 30 Myr and triangles represent member clusters with age between 30 and 80 Myr. There are 18 member clusters in OCC 1 whose age data are available. The $V_{GC}$ --- $R_{GC}$  diagram of OOC 1 indicates that 6 clusters with age less 30 Myr and 4 clusters with age great than 30 Myr distribute in a small region with $ 212 < V_{GC} <219 $ km s$^{-1}$ and $ 8.0 < R_{GC} < 8.5$ kpc, those clusters can be considered as  kinematic members of OCC 1. But in the $z_{max}$ --- $e$ diagram of OCC 1, it is difficult to identify any clustering of those member clusters. \textbf{We could only determine the orbits of 50\% of the assumed members of OCC 2.} Most of member clusters of OCC 2 have age between 200 and 400 Myr. The panels of OCC 2 indicate that we cannot detect any clustering of  the member clusters in the $V_{GC}$ --- $R_{GC}$ and $z_{max}$ --- $e$ diagrams. The OCC 3 and 4 only include few of member clusters, the panels of those two OCCs indicate no clustering of the member clusters can be identified. Fig.~\ref{occs} indicates that only young OCCs can be identified in the $V_{GC}$ --- $R_{GC}$  diagram, and it is difficult to detect clustering in OCs with the $z_{max}$ --- $e$ diagram.

More recently, using the radial velocity, proper motion, inclination, and Galactic latitude of 341 OCs with age less than 100 Myr and within 2.5 kpc from the Sun, \citet{dl08} studied the clustering in those clusters. Most of member clusters in the closest OCC detected by \citet{dl08} have age less than 30 Myr, which is consistent with that we have found in the $V_{GC}$ --- $R_{GC}$ diagram of OCC 1. \citet{dl08} pointed out that 20 Myr is the characteristic timescale of an OCC, after this timescale, the complex may no longer be recognizable in the space of the orbital elements as the majority of its member have evaporated; only in the context of corotation resonances or mergers, the nongenetically related dynamical groups of old OCs can be detected in the Galactic disc.

\section{summary and conclusions}
We have presented a sample of 488 OCs whose distances, radial velocities, and absolute proper motions are used to derive their orbits in the Galaxy. The kinematical and the orbital characteristics of this sample are analyzed. The main results are listed as following:

\begin{itemize}
\item For OCs  with errors in the spatial velocities less than 20 km s$^{-1}$ in our present sample,  the velocity ellipsoids are derived as ($\sigma_{U}$, $\sigma_{V}$, $\sigma_{W})=$(28.7,  15.8,  11.0) km s $^{-1}$. The ages for most of clusters in our present sample are less than 500 Myr, and this sample represents a young thin disc population. The velocity dispersions of OCs in the three velocity components increase with the age of the cluster subsample, which indicates the continuous heating of the disc (N04).

\item The orbits of OCs are calculated with three Galactic gravitational potential models. Considering the intrinsic variation of orbital parameters and the effects of observational uncertainties, the errors for the derived orbital parameters are determined with a Monte Carlo fashion.The major errors come from the observational uncertainties of the input data and are mainly affected by the errors in the distance data. The observational uncertainties mainly affect the derived orbital eccentricities $e$ and the maximum distances above the Galactic disc $z_{\max}$.

\item The total masses within given distance range for different Galactic models can affect the derived orbital parameter $R_{a}$ and the orbital period $T_{p}$. The disc models for different Galactic models mainly affect the vertical movement of clusters and change the derived $z_{\max}$ and $T_{z}$. The uncertainties in the derived orbital parameters caused by different models are smaller than those caused by observational errors in the input data.

\item The detection probability for a cluster at the given galactocentric distance is calculated and the largest detection probability is at the cluster's apogalacticon. For most of OCs in our present sample, their present observed positions are very close to their apogalacticons. The mean of orbital period for OCs is  about seven times longer than the time for clusters crossing the Galactic plane.

\item Based on the presently observed galactocentric distances, the radial metallicity gradient for clusters with $R_{\textrm{GC}} <13.5$ kpc is derived with a slope $-0.070\pm0.011$ dex kpc$^{-1}$, which is consistent with the previous studies \citep{fr02,ch03}. The radial metallicity gradient derived based on apogalactic distances for the same sample is $-0.082\pm0.014$ dex kpc$^{-1}$.

\item The orbital eccentricities $e$ for different populations: globular clusters, disc giants, disc F and G dwarf stars, and OCs are compared. The orbital eccentricities for globular clusters occupy a large range and with a mean of $\sim 0.5$. The mean of orbital eccentricities for the thick disc giants is $\sim 0.4$ which is bigger than the mean of $\sim 0.1$ for the thin disc giants, F and G dwarfs, and OCs. There are about 3.7\% clusters in our sample belonging to the thick disc.

\item \textbf{Using the $V_{GC}$ --- $R_{GC}$ diagram, only one OCC could be identified and most of its members are younger than 30 Myr. We find it difficult to detect any OCCs in the $z_{max}$ --- $e$ diagram.}
\end{itemize}

\section*{Acknowledgments}
We are grateful to an anonymous referee for a particularly constructive and prompt report. This work has been supported in part by the National Natural Science Foundation of China, No.10633020, 10603006, 10778720, 10873016,  and 10803007; and by National Basic Research Program of China (973 Program) No. 2007CB815403. Z.-Y. W. acknowledges support from the Knowledge Innovation Program of the Chinese Academy of Sciences. This research has made use of the WEBDA database, operated at the Institute for Astronomy of the University of Vienna.


\begin{thebibliography}{}
\bibitem[\protect\citeauthoryear{Alessi, Moitinho \& Dias}{Alessi et al.}{2003}]{al03} Alessi B. S., Moitinho A., Dias W. S., 2003, A\&A, 410, 565
\bibitem[\protect\citeauthoryear{Allen \& Martos}{1988}]{as88} Allen C., Martos M. A., 1988, Rev. Mexicana Astron. Astrof., 16, 25 
\bibitem[\protect\citeauthoryear{Allen \& Santill\'{a}n}{1991}]{as91} Allen C., Santill\'{a}n A., 1991, Rev. Mexicana Astron. Astrof., 22, 255
\bibitem[\protect\citeauthoryear{Allen, Moreno \& Pichardo}{Allen et al.}{2006}]{al06} Allen C., Moreno E., Pichardo B., 2006, ApJ, 652, 1150
\bibitem[\protect\citeauthoryear{Allen, Moreno \& Pichardo}{Allen et al.}{2008}]{al08} Allen C., Moreno E., Pichardo B., 2008, ApJ, 674, 237
\bibitem[\protect\citeauthoryear{Barbier-Brossat \& Figon}{2000}]{bb00}Barbier-Brossat M., Figon P., 2000, A\&AS, 142, 217
\bibitem[\protect\citeauthoryear{Barkhatova, Kutuzov \& Osipkov}{Barkhatova et al.}{1987}]{ba87} Barkhatova K. A., Kutuzov S. A., Osipkov L. P., 1987, Soviet Astronomy, 31, 501
\bibitem[\protect\citeauthoryear{Barkhatova, Osipkov \& Kutuzov}{Batkhatova et al.}{1989}]{ba89} Barkhatova K. A., Osipkov L. P., Kutuzov S. A., 1989, Soviet Astronomy, 33, 596
\bibitem[\protect\citeauthoryear{Baumgardt, Dettbarn \& Wielen}{Baumgardt et al.}{2000}]{ba00} Baumgardt H., Dettbarn C., Wielen R., 2000, A\&AS, 146, 251
\bibitem[\protect\citeauthoryear{Bergond, Leon \& Guibert}{Bergond et al.}{2001}]{be01} Bergond G., Leon S., Guibert J., 2001, A\&A, 377, 462
\bibitem[\protect\citeauthoryear{Bonatto et al.}{2006}]{bo06} Bonatto C., Kerber L. O., Bica E., Santiago B. X., 2006, A\&A, 446, 121
\bibitem[\protect\citeauthoryear{Bovy, Hogg \& Roweis}{Bovy et al.}{2009}]{bo09} Bovy J., Hogg D. W., Roweis S. T., 2009, ArXiv:0905.2980 [astro-ph]
\bibitem[\protect\citeauthoryear{Carraro \& Chiosi}{1994}]{ca94} Carraro G., Chiosi C., 1994, A\&A, 288, 751
\bibitem[\protect\citeauthoryear{Chen, Hou \& Wang}{Chen et al.}{2003}]{ch03} Chen L., Hou J. L.,  Wang J. J., 2003, AJ, 125, 1397
\bibitem[\protect\citeauthoryear{Chen et al.}{2007}]{ch07} Chen L., Hou J. L., Zhao J. L., de Grijs R.,  2007, in Jin W. J., Platais I., Perryman M. A. C., eds, Proc. IAU Symp. 248,  A Giant Step: from Milli- to Micro-arcsecond Astrometry. Kluwer, Dordrecht, p. 433
\bibitem[\protect\citeauthoryear{Dehnen \& Binney}{1998}]{de98} Dehnen W., Binney J., 1998, MNRAS, 298, 387
\bibitem[\protect\citeauthoryear{de la Fuente Marcos \& de la Fuente Marcos}{2008}]{dl08} de la  Fuente Marcos R., de la Fuente Marcos C., 2008, ApJ, 672, 342
\bibitem[\protect\citeauthoryear{de Oliveira et al.}{2002}]{de02} de Oliveira M. R., Fausti A., Bica E., Dottori H., 2002, A\&A, 390, 103
\bibitem[\protect\citeauthoryear{Dias, L\'{e}pine \& Alessi}{Dias et al.}{2001}]{di01} Dias W. S., L\'{e}pine J. R. D., Alessi B. S., 2001, A\&A, 376, 441
\bibitem[\protect\citeauthoryear{Dias, L\'{e}pine \& Alessi}{Dias et al.}{2002}]{di02p} Dias W. S., L\'{e}pine J. R. D., Alessi B. S., 2002, A\&A, 388, 168
\bibitem[\protect\citeauthoryear{Dias et al.}{2002}]{di02} Dias W. S., Alessi B. S., Moitinho A., L\'{e}pine J. R. D., 2002, A\&A, 389, 871
\bibitem[\protect\citeauthoryear{Dias \& L\'{e}pine}{2005}]{di05}  Dias W. S., L\'{e}pine J. R. D., 2005, ApJ, 629, 825
\bibitem[\protect\citeauthoryear{Dias et al.}{2006}]{di06} Dias W. S., Assafin M., Fl\'{o}rio V., Alessi B. S., L\'{i}bero V., 2006, A\&A, 446, 949
\bibitem[\protect\citeauthoryear{Dinescu, Girard \& van Altena}{Dinescu et al}{1999}]{di99} Dinescu D. I., Girard T. M., van Altena W. F., 1999, AJ, 117, 1792
\bibitem[\protect\citeauthoryear{Efremov}{1978}]{ef78} Efremov Y. N., 1978, Soviet Astronomy Letters, 4, 66
\bibitem[\protect\citeauthoryear{\'{E}igenson \& Yatsyk}{1988}]{ei88} \'{E}igenson A. M., Yatsyk O. S., 1988, Soviet Astronomy, 32, 168
\bibitem[\protect\citeauthoryear{ESA}{1997}]{esa} ESA, 1997, The Hipparcos and Tycho Catalogues, ESA SP-1200
\bibitem[\protect\citeauthoryear{Finlay et al.}{1995}]{fi95}Finlay J., Noriega-Crespo A., Friel E. D., Cudworth K., 1995, Bull. Am. Astron. Soc., 27, 1437
\bibitem[\protect\citeauthoryear{Flynn, Sommer-Larsen \& Christensen}{Flynn et al.}{1996}]{fsc96} Flynn C., Sommer-Larsen J., Christensen P. R., 1996, MNRAS, 281, 1027
\bibitem[\protect\citeauthoryear{Friel}{1995}]{fr95} Friel E. D., 1995, ARA\&A, 33, 381
\bibitem[\protect\citeauthoryear{Friel}{1999}]{fr99} Friel E. D., 1999, Ap\&SS, 265, 271
\bibitem[\protect\citeauthoryear{Friel et al.}{2002}]{fr02} Friel E. D., et al., 2002, AJ, 124, 2693
\bibitem[\protect\citeauthoryear{Frinchaboy \& Majewski}{2008}]{fr08}Frinchaboy P. M., Majewski S. R., 2008, ApJ, 136, 118
\bibitem[\protect\citeauthoryear{Fu et al.}{2009}]{fu09} Fu J., Hou J. L., Yin J., Chang R. X., 2009, ApJ, 696, 668
\bibitem[\protect\citeauthoryear{Holmberg, Nordstr\"{o}m \& Andersen}{Holmberg et al.}{2007}]{ho07} Holmberg J., Nordstr\"{o}m B., Andersen J., 2007, A\&A, 475, 519
\bibitem[\protect\citeauthoryear{H{\o}g et al.}{2000}]{ho00} H{\o}g E., et al., 2000, A\&A, 355, L27
\bibitem[\protect\citeauthoryear{Janes}{1979}]{j79} Janes K. A., 1979, ApJS, 39, 135
\bibitem[\protect\citeauthoryear{Janes, Tilley \& Lyng\r{a}}{Janes et al.}{1988}]{j88} Janes K. A., Tilley C.,  Lyng\r{a} G., 1988, AJ, 95, 771
\bibitem[\protect\citeauthoryear{Johnson \& Soderblom}{1987}]{jo87} Johnson D. R. H., Soderblom D. R., 1987, AJ, 93, 864
\bibitem[\protect\citeauthoryear{J{\o}rgensen \& Lindegren}{2005}]{jl05} J{\o}rgensen B. R.,  Lindegren L., 2005, A\&A, 436, 127
\bibitem[\protect\citeauthoryear{Keenan, Innanen \& House}{Keenan et al.}{1973}]{ke73} Keenan D. W., Innanen K. A., House F. C., 1973, AJ, 78, 173
\bibitem[\protect\citeauthoryear{Kerr \& Lynden-Bell}{1986}]{ke86} Kerr F. J., Lynden-Bell D., 1986, MNRAS, 221, 1023
\bibitem[\protect\citeauthoryear{Kharchenko}{2001}]{kh01} Kharchenko N. V., 2001, Kinematics and Physics of Celestial Bodies, 17, 409
\bibitem[\protect\citeauthoryear{Kharchenko, Pakulyak \& Piskunov}{Kharchenko et al.}{2003}]{kh03} Kharchenko N. V., Pakulyak L. K., Piskunov A. E., 2003, Astron. Rep., 47, 263
\bibitem[\protect\citeauthoryear{Kharchenko et al.}{2005}]{kh05}Kharchenko N. V., Piskunov A. E., R\"{o}ser S., Schilbach E., Scholz R. -D., 2005, A\&A, 438, 1163
\bibitem[\protect\citeauthoryear{Kharchenko \& Piskunov}{2006}]{kh06}Kharchenko N. V., Piskunov A. E., 2006, Astron. Astrophys. Trans. 25, 177
\bibitem[\protect\citeauthoryear{Kharchenko et al.}{2007}]{kh07} Kharchenko N. V., Scholz R. -D., Piskunov A. E., R\"{o}ser S., Schilbach E., 2007, Astron. Notes, 328, 889
\bibitem[\protect\citeauthoryear{Kovalevsky et al.}{1997}]{ko97} Kovalevsky J., et al., 1997, A\&A, 323, 620
\bibitem[\protect\citeauthoryear{Lada \& Lada}{2003}]{la03} Lada C. J., Lada E. A., 2003, ARA\&A, 41, 57
\bibitem[\protect\citeauthoryear{L\'{e}pine, Dias \& Mishurov}{L\'{e}pine et al.}{2008}]{le08}  L\'{e}pine J. R. D., Dias W. S.,  Mishurov Y., 2008, MNRAS, 386, 2081
\bibitem[\protect\citeauthoryear{Loktin \& Beshenov}{2003}]{lo03} Loktin A. V., Beshenov G. V., 2003, Astron. Rep., 47, 6
\bibitem[\protect\citeauthoryear{Lyng\r{a}}{1982}]{ly82} Lyng\r{a} G., 1982, A\&A, 109, 213
\bibitem[\protect\citeauthoryear{Lyng\r{a}}{1987}]{ly87a} Lyng\r{a} G., 1987, Catalogue of open cluster data, 5th edition, CDS, Strasbourg
\bibitem[\protect\citeauthoryear{Lyng\r{a} \& Palou\v{s}}{1987}]{ly87b} Lyng\r{a} G., Palou\v{s} J., 1987, A\&A, 188, 35
\bibitem[\protect\citeauthoryear{Magrini et al.}{2009}]{ma09} Magrini L., Sestito P., Randich S., Galli D., 2009, A\&A, 494, 95
\bibitem[\protect\citeauthoryear{Mermilliod, Mayor \& Udry}{Mermilliod et al.}{2008}]{me08}Mermilliod J. C., Mayor M., Udry S., 2008, A\&A, 485, 303,
\bibitem[\protect\citeauthoryear{Miyamoto \& Nagai}{1975}]{mn75} Miyamoto M., Nagai R., 1975, PASJ, 27, 533
\bibitem[\protect\citeauthoryear{Nordstr\"{o}m et al.}{2004}]{no04} Nordstr\"{o}m B., et al., 2004, A\&A, 418, 989
\bibitem[\protect\citeauthoryear{Odenkirchen \& Brosche}{1992}]{od92} Odenkirchen M., Brosche  P., 1992, Astron. Nachr., 313, 69
\bibitem[\protect\citeauthoryear{Odenkirchen et al.}{1997}]{od97} Odenkirchen M., Brosche  P., Geffert M., Tucholke H. -J., 1997, New. Astron., 2, 477
\bibitem[\protect\citeauthoryear{Paczy\'{n}ski}{1990}]{p90} Paczy\'{n}ski B., 1990, ApJ, 348, 485
\bibitem[\protect\citeauthoryear{Paunzen \& Netopil}{2006}]{pa06} Paunzen E., Netopil M., 2006, MNRAS, 371, 1641
\bibitem[\protect\citeauthoryear{Piskunov et al.}{2006}]{pi06} Piskunov A. E., Kharchenko N. V., R\"{o}ser S., Schilbach E., Scholz R. -D., 2006, A\&A, 445, 545
\bibitem[\protect\citeauthoryear{Platais, Kozhurina-Platais \& van Leeuwen}{Platais et al.}{1998}]{pl98} Platais I., Kozhurina-Platais V., van Leeuwen F., 1998, AJ, 116, 2423
\bibitem[\protect\citeauthoryear[{Plummer}{1911}]{pl11}Plummer H. C., 1911, MNRAS, 71, 460
\bibitem[\protect\citeauthoryear{Press et al.}{1992}]{pr92} Press W. H., Teukolsky S. A., Vetterling W. T. Flannery B. P., 1992, Numerical Recipes in Fortran: The Art of Scientific Computing, 2rd edn. Cambridge Univ. Press, Cambridge
\bibitem[\protect\citeauthoryear{Reid}{1993}]{re93} Reid M. J., 1993, ARA\&A, 31,345
\bibitem[\protect\citeauthoryear{R\"{o}ser et al.}{2007}]{ro07} R\"{o}ser S., Kharchenko N. V., Piskunov A. E., Schilbach E., Scholz R.-D., 2007, in Vesperini E., Giersz M., Sills A., eds, Proc. IAU Symp. 246, Dynamical Evolution of Dense Stellar Systems. Kluwer, Dordrecht, p. 115
\bibitem[\protect\citeauthoryear{Saio \& Yoshii}{1979}]{sa79} Saio H., Yoshii Y., 1979, PASP, 91, 553
\bibitem[\protect\citeauthoryear{Seabroke \& Gilmore}{2007}]{se07} Seabroke G. M., Gilmore G., 2007, MNRAS, 380, 1348
\bibitem[\protect\citeauthoryear{Soubiran, Odenkirchen \& Le Campion}{Soubiran et al.}{2000}]{so00} Soubiran C., Odenkirchen M., Le Campion J.-F., 2000, A\&A, 357, 484
\bibitem[\protect\citeauthoryear{Soubiran et al.}{2008}]{so08} Soubiran C., Bienaym\'{e} O., Mishenina T. V., Kovtyukh V. V., 2008,  A\&A, 480, 91
\bibitem[\protect\citeauthoryear{Twarog, Ashma \& Anthony-Twarog}{Twarog et al.}{1997}]{tw97}Twarog B. A., Ashma K. M., Anthony-Twarog B. J., 1997, AJ, 114, 2556
\bibitem[\protect\citeauthoryear{Urban, Wycoff \& Makarov}{2000}]{ur00} Urban S. E., Wycoff G. L., Makarov V. V., 2000, AJ, 120, 501
\bibitem[\protect\citeauthoryear{van Leeuwen}{2008}]{vl07a} van Leeuwen F., 2008, Hipparcos, the new reduction of the raw data, ASSL, Springer
\bibitem[\protect\citeauthoryear{van Leeuwen}{2007}]{vl07b} van Leeuwen F., 2007, A\&A, 474, 653
\bibitem[\protect\citeauthoryear{Wu et al.}{2002}]{wu02} Wu Z. Y., Tian K. P., Balaguer-N\'{u}\~{n}ez L., Jordi C., Zhao L., Guibert J., 2002, A\&A, 381, 464
\bibitem[\protect\citeauthoryear{Zacharias et al.}{2004}]{za04} Zacharias N., et al., 2004, AJ, 127, 3043
\end{thebibliography}
\end{document}